\documentstyle[10pt,aasms4,flushrt]{article}

\begin{document}
\normalbaselineskip=12pt
\normalbaselines

%

\title{A Convenient Set of Comoving Cosmological Variables
and Their Application}

\author{Hugo Martel\altaffilmark{1} and
        Paul R. Shapiro\altaffilmark{1}}

\altaffiltext{1}{Department of Astronomy, University of Texas, Austin,
                 TX 78712, USA}

\let \uth=\varepsilon

\def \gbf   {{\bf g}}
\def \kbf   {{\bf k}}
\def \pbf   {{\bf p}}
\def \qbf   {{\bf q}}
\def \rbf   {{\bf r}}
\def \ubf   {{\bf u}}
\def \vbf   {{\bf v}}
\def \xbf   {{\bf x}}
\def \Bbf   {{\bf B}}
\def \Sbf   {{\bf S}}
\def \Tbf   {{\bf T}}

\def \trho  {\tilde\rho}              \def \grad      {\nabla}
\def \tbrho {\tilde{\bar\rho}}        \def \divg      {\nabla\cdot}
\def \tr    {\tilde\rbf}              \def \laplacian {\nabla^2}
\def \trr   {\tilde r}                \def \curl      {\nabla\times}
\def \tuth  {\tilde\uth}
\def \ttime {\tilde t}
\def \tv    {\tilde\vbf}
\def \tvv   {\tilde v}
\def \tf    {\tilde\fbf}
\def \tphi  {\tilde\phi}
\def \tp    {\tilde p}
\def \tx    {\tilde x}
\def \tq    {\tilde q}
\def \th    {\tilde h}
\def \tgrad {\tilde\grad}
\def \tcs   {\tilde c_s}
\def \txi   {\tilde\xi}
\def \tk    {\tilde k}
\def \tpbf  {\tilde\pbf}
\def \tqbf  {\tilde\qbf}

\def \omegabf  {\mbox{\boldmath$\omega$}}
\def \tomega   {\tilde{\omegabf}}

\def \Omo   {\Omega_0}
\def \Omx   {\Omega_{X0}}

\def\ast{\mathchar"2203} \mathcode`*="002A      

%

\begin{abstract}
A set of cosmological variables, 
which we shall refer to as ``supercomoving variables,'' are presented
which are an alternative to the standard comoving variables,
particularly useful for describing the gas dynamics of cosmic structure
formation. For ideal gas with a ratio of specific heats $\gamma=5/3$,
the supercomoving position, velocity, and thermodynamic properties
(i.e. density, temperature, and pressure) of matter are constant in
time in a uniform, isotropic, adiabatically expanding universe.
Expressed in terms of these supercomoving variables,
the nonrelativistic, cosmological fluid conservation equations of the
Newtonian approximation and the Poisson equation
closely resemble their noncosmological counterparts.
This makes it 
possible to generalize noncosmological results and techniques 
to address problems involving departures from uniform, adiabatic Hubble
expansion in a straightforward way, for a wide range of cosmological models.
These variables were initially introduced by Shandarin (1980) to describe 
structure formation in matter-dominated models. In this paper, we 
generalize supercomoving variables to models with a uniform contribution to the
energy density corresponding to a nonzero cosmological constant,
domain walls, cosmic strings, a nonclumping form of
nonrelativistic matter (e.g. massive neutrinos in the presence
of primordial density fluctuations of small wavelength), or a radiation 
background. Each model is characterized by the value of the density 
parameter~$\Omo$ of the non-relativistic matter component in which density 
fluctuation is possible, and the density parameter~$\Omx$ of the additional, 
nonclumping component. For each type of nonclumping background, we identify 
{\it families} within which different values of $\Omo$ and $\Omx$ lead to
fluid equations and solutions in supercomoving
variables which are independent of the cosmological parameters~$\Omo$ 
and~$\Omx$. We also generalize the description to include the effects of
nonadiabatic processes such as heating, radiative cooling, thermal conduction
and viscosity, as well as magnetic fields in the MHD approximation.

As an illustration, we describe three familiar cosmological problems in
supercomoving variables:
the growth of linear density fluctuations, the nonlinear collapse
of a one-dimensional plane-wave density fluctuation
leading to pancake formation, and the well-known Zel'dovich approximation for 
extrapolating the linear growth of density fluctuations in three dimensions
to the nonlinear stage. 

\end{abstract}


\newpage

%

\section{INTRODUCTION}

The modern search for an explanation for the origin of the observed structure
in the universe has led to the need to solve multi-scale, highly nonlinear
problems of great complexity which involve the coupling of gravitational
dynamics, gas dynamics, and cosmological expansion. In what follows, we
shall describe a useful tool -- a change of variables --- which simplifies the 
application of the more familiar results and techniques developed to describe
a noncosmological gas in a nonexpanding background to problems in cosmology.

The description of the formation and evolution of large-scale structure
in the universe is greatly simplified if we restrict our attention
to systems with length scales much larger than
the Schwarzschild radius of the largest mass concentrations in the universe 
and much smaller than the horizon, $c/H$, where $H$ is the Hubble constant,
with nonrelativistic peculiar motions and temperatures. In this limit,
known as the Newtonian approximation, we can use the ordinary 
noncosmological, nonrelativistic fluid conservation
equations and the Poisson equation,
modified to include the effects of universal, adiabatic, Hubble
expansion, to describe the evolution of the baryonic and dark matter that
form the galaxies and large-scale structure we observe today. 

A straightforward change of variables is sometimes introduced in which
the uniform, isotropic Hubble expansion is factored out so as
to obtain a set of
equations that describes the evolution of departures from a uniform,
isotropic, structureless universe. This is achieved by rewriting the
fluid conservation
equations in~{\it comoving variables}. These variables are defined by
\begin{equation}
\xbf={\rbf\over a(t)}\,,
\end{equation}

\noindent and
\begin{equation}
\rho_{\rm co-mov}=\rho a(t)^3\,,
\end{equation}

\noindent where $\rbf$ and $\rho$ are the proper distance and the
mass density, respectively, $\xbf$ is the comoving position,
$\rho_{\rm co-mov}$ is the comoving density, and
$a(t)$ is the Robertson-Walker scale factor, whose 
time-evolution is described by the Friedmann equation (see eq.~[4] below).
In these comoving variables, in the absence of structure, the mass points
are at rest and the density of ordinary, nonrelativistic matter
is constant in time. For a detailed
description of the Newtonian dynamics in comoving variables, see
Weinberg (1972) and Peebles (1980, hereafter P80).

The fluid conservation equations
in comoving variables contain several
terms which distinguish them from their noncosmological counterparts.
In particular,
the momentum and energy equations contain ``drag terms'' that make the
peculiar velocities and the thermal energies decay with time, even in
the absence of departures from uniformity involving pressure or gravitational
potential gradients to act as driving forces. 
This reflects the fact that universal expansion, itself, causes adiabatic
cooling (i.e. $PdV$ work is done), while the kinematics of universal 
expansion causes a test particle which moves relative to the universal
expansion, with no peculiar forces acting on it, to appear to fall behind
relative to the expanding coordinate system.

A further change of variables was introduced by Shandarin (1980, hereafter
S80), referred to as ``tilde variables'' when modified and applied
by Shapiro \& collaborators (e.g. Shapiro, Struck-Marcell, \& Melott 1983;
Shapiro \& Struck-Marcell 1985; Shapiro et al. 1996;
Valinia et al. 1996). These new variables extended the concept of ``comoving''
to the limit in which the position, velocity, and thermodynamic properties
(i.e. density, temperature, and pressure) of matter in a uniform, isotropic, 
adiabatically expanding, matter-dominated Friedmann universe all remain
stationary. In addition, by replacing proper time by a new time variable, the
so-called ``drag terms'' mentioned above disappeared from the momentum equation
when expressed in terms of these ``tilde variable.'' For an ideal gas with
a ratio of specific heats $\gamma=5/3$, the ``drag term'' in the energy
equation disappeared, as well. For such a gas, the complete set of three fluid
conservation equations, for matter, momentum, and energy, 
when expressed in tilde variables,
are identical to the standard, noncosmological 
fluid equations. Henceforth, we shall refer to these tilde variables
as ``supercomoving variables.'' instead, in order to give a more descriptive
name.

The use of supercomoving variables makes it possible 
to take the known solutions of non-cosmological problems and 
generalize them in a straight-forward way
to cosmological situations. An example of this
is provided by Voit (1996), who used supercomoving variables to
address the problem of the evolution of intergalactic blast
waves in a matter-dominated universe. The similarity of the 
fluid equations in supercomoving and noncosmological variables
also means that analytical methods and numerical simulation
algorithms which are suitable for studying non-cosmological
problems can 
often be applied to cosmological ones with essentially no modification.
Examples of the latter include the 1D, Lagrangian hydrodynamics
method developed by Shapiro et al. (1983) and
Shapiro \& Struck-Marcell (1985) to study cosmological pancakes and the 
new anisotropic version of Smoothed Particle Hydrodynamics (SPH), called
Adaptive SPH (ASPH), developed recently by Shapiro et al. (1996) to
study both cosmological and noncosmological gas dynamics by numerical
simulation in 2D and 3D.

Shandarin (S80) also showed that, with an appropriate
normalization of the scale factor~$a(t)$, it was possible to eliminate
all dependences, implicit or explicit, upon 
the density parameter~$\Omo$. All matter-dominated
cosmological models can be grouped into three distinct
families, $\Omo<1$, $\Omo=1$, and $\Omo>1$, and within each family, the
fluid conservation equations and their solutions, and the solutions
of the Friedmann equation, are identical. There is no distinction
in supercomoving variables between a model with~$\Omo=0.2$ and~$\Omo=0.99$, 
for instance.

The supercomoving variables introduced by Shandarin (S80) were
restricted to matter-dominated models (i.e. models
in which the mean energy density is dominated by nonrelativistic matter
[baryonic and dark], so $\Omx=0$).
There are several reasons, however, for considering cosmological models
that contain more than just the nonrelativistic component.
For the particular case of a nonzero cosmological constant, these reasons
have been summarized by Carroll, Press, \&~Turner (1992),
Ostriker \&~Steinhardt (1995), and Krauss \& Turner (1995).
The case of a more general, smooth background component has been discussed
recently by Turner \&~White (1997). Some of the arguments are as follows.
Dynamical estimates of the density parameter~$\Omo$ 
have consistently given values smaller than unity
(see, for example, Peebles 1993 and references therein)
(see, however, Dekel et al. 1993, and references therein
for an exception to this trend),
in contradiction with the most common inflationary scenario 
(Guth 1981; Linde 1982; Albretch \& Steinhardt 1982) which requires~$\Omo=1$.
(We note, however, that inflationary models with $\Omo<1$ are now 
being considered [Lyth \&~Steward 1990; Ratra \&~Peebles 1994].)
To reconcile the dynamical estimates of $\Omo$ with standard inflation, we 
must postulate the existence of a second component which must be uniform
in space in order to escape detection by dynamical methods (hence, we are 
not referring to the dark matter component in galaxy clusters and
galactic halos; this
component is already included in dynamical estimates of~$\Omo$).
The presence of this uniform component such as domain
walls or a cosmological constant also increases the age of the universe
for a given value of~$H_0$. This helps to reconcile current estimates of~$H_0$,
which favor $H_0\approx60-80\,\rm km\,s^{-1}Mpc^{-1}$
(Riess, Kirshner, \&~Press 1995; Freedman 1996;
Giovanelli et al 1996),
with lower limits to the age of the universe based on
estimates of globular clusters ages, the well-known age problem. 
With the detection of temperature anisotropy by the COBE satellite DMR
experiment, it has become possible to fix the amplitude and constrain the
wavelength dependence of the primordial density fluctuations responsible for
large-scale structure (LSS) formation, at least at the superhorizon scale
at recombination to which the COBE measurement is sensitive (see Bunn \&~White
1997, and references therein). The much-studied, standard, flat, 
matter-dominated Cold Dark Matter (CDM) model for the origin of galaxies and 
LSS is, however, incompatible with these COBE constraints since it tends
to exaggerate the structure on the smaller scales measured by the statistical
clustering properties of galaxies and LSS in comparison with galaxy data 
(see Ostriker 1993 [\S3.2], and references therein;
Ostriker \& Steinhardt 1995, and references therein). The flat CDM model
can be reconciled with these COBE results, however, if there is a uniform
background component like a cosmological constant which dominates at late
times over the nonrelativistic matter component.

The main goal of this paper, therefore,
is to generalize the original supercomoving variables of Shandarin, which are
limited to matter-dominated models, to include
models containing an additional, uniform component. In the
process, we will demonstrate that different cosmological models corresponding
to different combinations of the fundamental cosmological parameters can be 
grouped into ``families'' in which the evolution of physical quantities
is the same when expressed in supercomoving variables.
We will follow Shapiro et al. (1983) and Shapiro \& Struck-Marcell (1985)
in the inclusion of non-adiabatic processes
like external heating, radiative cooling, and thermal conduction,
and extend their description to
include viscosity, vorticity, and the effects of magnetic fields in the MHD
approximation. 

The remainder of the paper is organized as follows. In
\S2, we present the fluid equations in noncosmologial variables.
In \S3, we define the supercomoving variables.
In \S4, we derive the supercomoving form of the fluid equations.
In \S5, we present the supercomoving form of the collisionless
Boltzman equation. 
In \S6, we derive the cosmic energy equation in supercomoving variables.
In \S7, we solve the Friedmann equation for the various models
considered in this paper, and derive the conditions required in order to have
families of solutions which are independent of the particular values of the
cosmological parameters for different combinations of these parameters.
In \S8, we present the solutions of the linear perturbation equations
for the various models considered. 
In \S9, we discuss exact, nonlinear solutions for the collapse of plane-wave
density fluctuations, prior to shock and caustic formation. 
In \S10, we present the supercomoving form of the Zel'dovich approximation.
Summary and conclusions are presented in \S11.
Our Appendix~A discusses additional physical processes, including viscosity,
heating, cooling, thermal conduction, vorticity, and magnetic fields in the
MHD approximation.

\section{BASIC EQUATIONS IN NONCOSMOLOGICAL COORDINATES}

\subsection{The Cosmological Background}

The evolution of the cosmological
background is determined by the relative contribution by
its constituent components to the mean mass-energy density of the universe, 
as it appears in the Einstein field equations.
For an isotropic, homogeneous universe with
a given composition, this evolution is expressed in terms of the 
Friedmann-Robertson-Walker metric with a cosmic scale 
factor~$a(t)$ whose time-evolution is given by the Friedmann equation.

In this paper, we shall
consider two-component models. The first component, called
the {\it nonrelativistic (NR) component}, comprises all forms of matter,
luminous or dark, baryonic or non-baryonic, that can cluster under 
the action of gravity,
and whose {\it mean} energy density varies with time as~$a(t)^{-3}$. 
The second component, which we shall refer to as the {\it X~component},
does not clump and, instead, has a
{\it uniform} energy density $\rho_X$ which varies with expansion
according to
\begin{equation}
\rho_X(t)\propto a(t)^{-n}\,,
\end{equation}

\noindent where~$n$ is a non-negative constant. This equation was first 
introduced by Fry (1985) (see also 
Charlton~\& Turner 1987; Silveira \&~Waga 1994; Martel 1995; 
Dodelson, Gates, \&~Turner 1996; Turner \&~White 1997).
By appropriate choice of~$n$, we recover
models with a nonzero cosmological constant ($n=0$), domain walls ($n=1$),
string networks ($n=2$), vacuum stress ($n=3$),
or a radiation background ($n=4$). In this paper, we
shall derive a form of the supercomoving variables applicable to all of
these various models. Notice that models with massive, nonrelativistic 
neutrinos are also described by equation~(3) with~$n=3$, as long as
the length scale of the problem studied is much shorter than the damping
scale of the neutrinos.
For $n\ne3$, the X~component has a relativistic pressure~$P_X=\nu\rho_Xc^2$,
where $c$ is the speed of light and $\nu=n/3-1$, which enters in the
Newtonian form of the Poisson equation as an additional source term
(see eq.~[8] below).

The time-evolution of the scale factor~$a(t)$ 
is described by the Friedmann equation.
For the two-component models considered in this paper, this 
equation takes the form
\begin{equation}
\biggl({\dot a\over a}\biggr)^2=H(t)^2=H_0^2\Biggl[
(1-\Omo-\Omx)\biggl({a\over a_0}\biggr)^{-2}
+\Omo\biggl({a\over a_0}\biggr)^{-3}+
\Omx\biggl({a\over a_0}\biggr)^{-n}\Biggr]\,,
\end{equation}

\noindent (Martel 1995)
where we use subscripts~0 to indicate the present value of 
time-dependent quantities. The first term in the right hand side is the 
curvature term $-k/a^2$, where $k=-H_0^2a_0^2(1-\Omo-\Omx)$. The other
two terms represent the nonrelativistic 
and X~component, respectively. The density 
parameters $\Omega$ and~$\Omega_X$ are defined by $\Omega=\bar\rho/\rho_c$
and~$\Omega_X=\rho_X/\rho_c$, where $\bar\rho$ and~$\rho_X=\bar\rho_X$ 
are the mean density
of the nonrelativistic and X~component, respectively, and 
$\rho_c\equiv3H^2/8\pi G$ is the critical density, defined as the mean
density of a flat universe, 
{\it in the absence of the X~component}.

\subsection{Fluid Conservation Equations in the Newtonian Approximation}

The dynamics of a self-gravitating fluid 
in the Newtonian approximation is described by the basic
hydrodynamical equations for the conservation of
mass, momentum, and energy,
coupled to the Poisson equation and the equation of state.
The Newtonian approximation and its validity in cosmology is
discussed by P80. It is valid in the
limit in which all length scales are much smaller than the horizon
size, $c/H(t)$, and much larger than the Schwarzschild radii of objects
within the region of study, and temperatures, equations of state, and peculiar
velocities are nonrelativistic.
In proper coordinates and in the absence of viscosity
or dissipative processes like thermal conduction or radiative cooling,
the continuity, momentum,
and energy equations, and the equation of state take the form
\begin{eqnarray}
&&{\partial\rho\over\partial t} +\divg(\rho\ubf)=0\,,\\
&&{\partial\ubf\over\partial t} 
+(\ubf\cdot\grad)\ubf=-\grad\Phi-{\grad p\over\rho}\,,\\
&&{\partial\uth\over\partial t}
+\ubf\cdot\grad\uth=-{p\over\rho}\divg\ubf\,,\\
&&p=(\gamma-1)\uth\rho\,,
\end{eqnarray}

\noindent where $\rho$ is the matter
density, $\uth$ is the specific internal energy, $p$ is the gas
pressure, $\ubf$ is the velocity, $\Phi$ is the gravitational potential,
and $\gamma$ is the ratio of specific heats.
These equations also assume that there is no interaction besides gravity
between the nonrelativistic component and the X~component.
In Appendix~A, we generalize these equations to include additional
physical processes, such as heating, cooling, thermal conduction,
viscosity, and magnetic fields.

\subsection{Gravitational Potential in the Newtonian 
Approximation: Modified Poisson Equation}

The Newtonian form of the Poisson equation must be modified to account
for the presence of the X~component.
To derive the Poisson equation, we consider the acceleration 
$g^\alpha$ between two freely-falling particles separated by a 
proper distance $\xi^\alpha$. In the {\it small-scale limit}, where
$\xi^\alpha$ is much smaller than the radius of curvature of the universe,
the space-time is locally Minkowskian, and the equation for the
acceleration reduces to
\begin{equation}
g^\alpha=R^\alpha{}_{00\beta}\xi^\beta\,,
\end{equation}

\noindent (Peebles 1993, p.~268) where $R^i{}_{jkl}$ is the Riemann-Christoffel
curvature tensor, and where we sum over repeated indicies. 
By taking the divergence of equation~(9), we get
\begin{equation}
\laplacian\Phi\equiv-\divg\gbf=4\pi GT^{\alpha\alpha}\,,
\end{equation}

\noindent where $\Tbf$ is the stress-energy tensor.
For the models we consider, the total stress-energy tensor 
in the fluid rest frame is given by
\begin{equation}
\Tbf=\left[
       \matrix{\rho+\rho_X & 0           & 0           & 0 \cr
               0           & p_X/c^2     & 0           & 0 \cr
               0           & 0           & p_X/c^2     & 0 \cr
               0           & 0           & 0           & p_X/c^2 \cr}\right]
       \,,
\end{equation}

\smallskip

\noindent where $\rho_X$ and $p_X$ are the energy density and pressure
of the X~component, respectively, and $c$ is the speed of light. 
In equation~(11), we neglected the nonrelativistic pressure term $p/c^2$,
which is smaller than $\rho$ by a factor of order $(c_s/c)^2$, where $c_s$ is
the sound speed.
The evolution of the density $\rho_X$ of the X~component is given by
\begin{equation}
{d\over da}(\rho_Xa^3)=-{3p_Xa^2\over c^2}
\end{equation}

\noindent (Weinberg 1972, p.~472). 
By combining equations~(3) and~(12), then yield
\begin{equation}
{p_X\over c^2}=\biggl({n\over3}-1\biggr)\rho_X\,.
\end{equation}

\noindent Equations~(10), (11), and~(13) reduces to
\begin{equation}
\laplacian\Phi=4\pi G\Big[\rho+(n-2)\rho_X\Big]\,.
\end{equation}

\noindent This is the final form of the Poisson equation in the presence
of the X~component. Notice that the contribution of the X~component to the
gravitational source term is negative for $n<2$. This implies that the 
gravitational force resulting from the presence of domain walls or a 
cosmological constant is repulsive.

\section{SUPERCOMOVING VARIABLES}

One of the most desirable features of the comoving and supercomoving variables
is that in these variables, in the absence of structure, the density
is constant in time and mass elements are at rest. 
In supercomoving variables, a gas with ratio of specific heats $\gamma=5/3$
also has its thermodynamic variables remain constant in the absence of 
structure. We preserve these two properties when we generalize supercomoving
variables to the cases involving an additional nonclumping background 
component. by making the following definitions,
\begin{eqnarray}
\tr&=&{\rbf\over ar_\ast}\,,\\
\trho&=&{a^3\rho\over\rho_\ast}\,,\\
\tv&=&{a\vbf\over v_\ast}\,,\\
d\ttime&=&{dt\over a^2t_\ast}\,,\\
\tphi&=&{a^2\phi\over\phi_\ast}\,,\\
\tp&=&{a^5p\over p_\ast}\,,\\
\tuth&=&{a^2\uth\over\uth_\ast}\,,
\end{eqnarray}

\noindent where the peculiar velocity $\vbf$ is defined by
\begin{equation}
\vbf=\ubf-H\rbf\,,
\end{equation}

\noindent
where $H\equiv a^{-1}da/dt$ is the Hubble constant,
and the peculiar gravitational potential $\phi$ is
related to the Eulerian gravitational potential $\Phi$ by
\begin{equation}
\Phi={2\pi G\bar\rho r^2\over3}+{2(n-2)\pi G\bar\rho_X r^2\over3}+\phi
={\Omo H_0^2\over4}\biggl({a_0\over a}\biggr)^3r^2
 +{\Omx H_0^2(n-2)\over4}\biggl({a_0\over a}\biggr)^nr^2+\phi
\,.
\end{equation}

\noindent This is a standard transformation, except for the second term
which we introduce to take into account the presence of the
X~component. In equations~(15)--(21), the quantities
$r_\ast$, $\rho_\ast$,
$\vbf_\ast$, $t_\ast$, $\phi_\ast$, $p_\ast$, and $\uth_\ast$
are fiducial values. Only $r_\ast$ is independent. The other fiducial 
quantities are defined by
\begin{eqnarray}
\rho_\ast&\equiv&\bar\rho_0={3H_0^2\Omo\over8\pi G}\,,\\
t_\ast&\equiv&{2\over H_0}\biggl({f_n\over\Omo a_0^3}\biggr)^{1/2}\,,\\
v_\ast&\equiv&{r_\ast\over t_\ast}\,,\\
\phi_\ast&\equiv&r_\ast^2/t_\ast^2=v_\ast^2\,,\\
p_\ast&\equiv&\rho_\ast r_\ast^2/t_\ast^2=\rho_\ast v_\ast^2\,,\\
\uth_\ast&\equiv&p_\ast/\rho_\ast=v_\ast^2\,,
\end{eqnarray}

\noindent where $\bar\rho_0$ is the cosmic mean matter density at
the present epoch and the parameter $f_n$ is defined by
\begin{equation}
f_n=\cases{1\,,&$n\neq3$;\cr
\displaystyle{\Omo\over\Omo+\Omx}\,,&$n=3$.\cr}
\end{equation}

For any particular cosmological model, we can eliminate $a(t)$ using
the Friedmann equation, and integrate equation~(18) to get an explicit 
relation between~$t$ and~$\ttime$. For future reference, equation~(18),
together with the Friedman equation~(4),
implies the following expressions for the Hubble constant and the first and
second derivatives of the scale factor,
\begin{eqnarray}
H(t)&=&{1\over a^3t_\ast}{da\over d\ttime}
\equiv{1\over a^2t_\ast}{\cal H}\,,\\
\biggl({da\over d\ttime}\biggr)^2&=&
t_\ast^2H_0^2a_0^6\Biggl[
(1-\Omo-\Omx)\biggl({a\over a_0}\biggr)^4+
\Omo\biggl({a\over a_0}\biggr)^3+
\Omx\biggl({a\over a_0}\biggr)^{6-n}\Biggr]\,,\\
{d^2a\over d\ttime^2}&=&
t_\ast^2H_0^2a_0^5\Biggl[
2(1-\Omo-\Omx)\biggl({a\over a_0}\biggr)^3+
{3\Omo\over2}\biggl({a\over a_0}\biggr)^2+
\biggl(3-{n\over2}\biggr)
\Omx\biggl({a\over a_0}\biggr)^{5-n}\Biggr]\,,
\end{eqnarray}

\noindent where equation~(31) defines the quantity~${\cal H}=a^{-1}da/d\ttime$,
the supercomoving Hubble constant
By combining equations~(31) and~(32), we can eliminate
$da/d\ttime$ and express $\cal H$ as a function of~$a$.

\section{SUPERCOMOVING FLUID EQUATIONS}

\subsection{Transformation of Derivatives}

To convert the hydrodynamical equations~(5)--(8) and~(14) to supercomoving
variables, we need to express the derivatives relative to~$\rbf$
at fixed~$t$ and $t$ at fixed~$\rbf$ as functions of the derivatives
relative to~$\tr$ at fixed~$\ttime$ and~$\ttime$ at fixed~$\tr$,
using equations~(15) and~(18).
This yields
\begin{eqnarray}
\biggl({\partial f\over\partial t}\biggr)_\rbf
&=&{1\over a^2t_\ast}\Biggl[\biggl({\partial f\over\partial\ttime}\biggr)_{\tr}
-{\cal H}\tr\cdot(\tgrad f)_{\ttime}\Biggr]\,,\\
(\grad f)_t&=&{1\over ar_\ast}(\tgrad f)_{\ttime}\,,
\end{eqnarray}

\noindent where $\tgrad$ is the gradient relative to $\tr$. 

\subsection{The Continuity Equation}

To derive the form of the continuity equation in supercomoving variables,
we substitute equations~(15), (16), (17), (22), (24), (26),
(34), and~(35) into
equation~(5). After some algebra, we get
\begin{equation}
{\partial\trho\over\partial\ttime} +\tgrad\cdot(\trho\tv)=0\,.
\end{equation}

\noindent This equation has exactly the same form as equation~(5).

\subsection{The Momentum Equation}

We substitute equations~(15), (16), (17), (19), (20), (22)--(28),
(34), and~(35) into equation~(6),
and eliminate the terms in~$da/d\ttime$ and~$d^2a/d\ttime\,^2$
using equations~(32) and~(33). We get, after some algebra,
\begin{equation}
{\partial\tv\over\partial\ttime} +(\tv\cdot\tgrad)\tv=-{\tgrad\tp\over\trho}
-\tgrad\tphi\,.
\end{equation}

\noindent Again, we have obtained an equation that has the same form
in supercomoving variables as in noncosmological variables.
The corresponding equation in comoving variables contains an
additional drag term.

\subsection{The Poisson Equation}

We substitute equations~(16), (19), (23), (24), (25), (27), and~(35) 
into equation~(14), and get
\begin{equation}
\tgrad^2\tphi=6af_n\biggl({\trho\over\tbrho}-1\biggr)\,,
\end{equation}

\noindent where we used~$\tbrho\equiv a^3\bar\rho/\rho_\ast
=a^3\bar\rho/\bar\rho_0=a_0^3$ (see eqs.~[16] and~[23]) to eliminate
$a_0$ in equation~(24).
For $n=3$, we can rewrite equation~(38) as
\begin{equation}
\tgrad^2\tphi=
6a\biggl({\trho_{\rm tot}\over\tbrho_{\rm tot}}-1\biggr)\,,
\end{equation}

\noindent where $\rho_{\rm tot}=\rho+\bar\rho_X$ is the total density,
{\it including the X~component}. 
The reason for treating the case $n=3$ differently will be
explained in \S7.3.3 below.

\subsection{The Equation of State}

We substitute equations~(16), (20), (21), (24), (28), and~(29)
into equation~(8). The equation of state becomes
\begin{equation}
\tp=(\gamma-1)\trho\tuth\,,
\end{equation}

\noindent which has the same form as equation~(8).

\subsection{The Energy Equation}

We substitute equations~(15), (16), (17), (20), (21), (22), (24), (26), (28),
(29), (34), and~(35)
into equation~(7). The energy equation becomes
\begin{equation}
{\partial\tuth\over\partial\ttime} +\tv\cdot\tgrad\tuth+{\cal H}
(3\gamma-5)\tuth=-{\tp\over\trho}\tgrad\cdot\tv\,.
\end{equation}

\noindent This equation contains a drag term, which vanishes only for
$\gamma=5/3$. 
We can derive the evolution of the mean 
specific internal energy $\tilde{\bar\uth}$
by considering a universe with no fluctuation. 
In this case, $\tv=0$
and~$\partial/\partial\ttime=d/d\ttime$. Equation~(41) then becomes
\begin{equation}
{d\tilde{\bar\uth}\over d\ttime} +{\cal H}
(3\gamma-5)\tilde{\bar\uth}=0\,.
\end{equation}

\noindent The solution is
\begin{equation}
\tilde{\bar\uth}(t)=\tilde{\bar\uth}_0\biggl({a\over a_0}\biggr)^{5-3\gamma}
\,.
\end{equation}

\noindent Hence, the mean specific internal energy decreases with expansion
for $\gamma>5/3$, and increases for~$\gamma<5/3$. Since the mean mass density
does not vary with expansion in supercomoving
variables, the equation of state gives
\begin{equation}
\tilde{\bar p}(t)=\tilde{\bar p}_0\biggl({a\over a_0}\biggr)^{5-3\gamma}
\,.
\end{equation}

The reason for the absence of drag term in equation~(41) for
the case $\gamma=5/3$ only is easily
understood. The momentum equation~(37) contains no drag term. 
From this equation,
we can derive an equation for the evolution of the specific
kinetic energy~$\tilde v^2/2$, which contains no drag term either.
In the case of $\gamma=5/3$, the internal energy of the gas is nothing more
than the 
microscopic specific kinetic energy of the particles that compose the gas.
Hence, because of the absence of drag term, the mean 
specific internal energy remains 
constant as the universe expands. But if $\gamma$ is not 5/3, then there is 
more to the internal energy of the gas than just the specific kinetic energy 
of the particles. For instance, for diatomic gases with $\gamma=7/5$,
some of the internal energy is in the form of molecular rotation. Hence, the
drag term in the energy equation~(41) represents the component
of the internal energy that is not associated with the microscopic motion
of the atoms and molecules in the gas.

\section{THE SUPERCOMOVING COLLISIONLESS BOLTZMANN EQUATION}

The equations derived in \S4 describe the evolution of a collisional
gas. These equations cannot be used for describing the 
evolution of a collisionless system of particles, because the
velocity~$\ubf$ is a multivalued quantity. The proper way to describe 
such system is in terms of a distribution function~$f(\rbf,\ubf,t)$ which
is defined as the number density of particles in phase space. The evolution
of the distribution function is described by the 
collisionless Boltzmann equation,
also known as the Vlasov equation. In noncosmological 
variables, this equation takes the form
\begin{equation}
{\partial f\over\partial t}+\ubf\cdot\grad f-\grad\Phi\cdot\grad_\ubf f=0
\,,
\end{equation}

\noindent where the gravitational potential is obtained by solving the
Poisson equation, which, for collisionless systems, is given by
\begin{equation}
\laplacian\Phi=4\pi G\biggl[\int f\,d^3u+\rho_b\biggr]\,,
\end{equation}

\noindent where we included the baryon density $\rho_b$ to take into account
systems in which the nonrelativistic component is a mixture of a 
collisionless and a collisional (gaseous) component.
After switching to supercomoving variables, the Boltzmann equation becomes
\begin{equation}
{\partial \tilde f\over\partial \ttime}+\tv\cdot\tgrad\tilde f
-\tgrad\tphi\cdot\tgrad_{\tv}\tilde f=0\,,
\end{equation}

\noindent where $\tilde f(\tr,\tv,\ttime\,)\equiv f(\rbf,\ubf,t)$,
and the Poisson equation becomes
\begin{equation}
\laplacian\tphi=\cases{
6a\biggl({\int
\displaystyle\tilde f\,d^3\tv+\trho_b\over\displaystyle\langle\scriptstyle\int
\displaystyle\tilde f\,d^3\tv\rangle+\tilde{\bar\rho}_b}-1\biggr)
\,,&$n\neq3$;\cr
6a\biggl({\int
\displaystyle\tilde f\,d^3\tv+\trho_X+\trho_b\over\displaystyle\langle
\scriptstyle\int
\displaystyle\tilde f\,d^3\tv\rangle
+\trho_X+\tilde{\bar\rho}_b}-1\biggr)
\,,&$n=3$;\cr}
\end{equation}

\noindent where $\langle\,\rangle$ indicates space-averaging. 

\section{THE COSMIC ENERGY EQUATION}

Using the fluid equations derived in the previous sections, we can derive
an equation for the evolution of the total energy. Our derivation follows 
the same lines as the one presented in P80 for the cosmic energy
equation in comoving variables. Taking the dot product
of $\trho\tv$ with the momentum equation~(37), we get
\begin{equation}
{1\over2}{\partial(\trho\tvv^2)\over\partial\ttime}
-{\tvv^2\over2}\biggl[{\partial\trho\over\partial\ttime}
                      +\tgrad\cdot(\trho\tv)\biggr]
=-\tgrad\cdot\biggl(\tp\tv+\trho\tphi\tv+{\trho\tvv^2\tv\over2}\biggr)
+\tp\tgrad\cdot\tv
+\tphi\tgrad\cdot(\trho\tv)\,.
\end{equation}

\noindent Using the continuity equation~(36), we cancel the term in square
brackets, and eliminate $\tgrad\cdot(\trho\tv)$ in the last term.
We also use the energy equation~(41) to eliminate the $\tp\tgrad\cdot\tv$ term.
After some additional vector manipulation, we get
\begin{equation}
{1\over2}{\partial(\trho\tvv^2)\over\partial\ttime}
=-\tgrad\cdot\biggl(\tp\tv+\trho\tphi\tv+{\trho\tvv^2\tv\over2}
                  +\trho\tuth\tv\biggr)
-{\partial(\trho\tuth)\over\partial\ttime}
+\tuth\biggl[{\partial\trho\over\partial\ttime}
                      +\tgrad\cdot(\trho\tv)\biggr]
-{\cal H}(3\gamma-5)\trho\tuth
-\tphi{\partial\trho\over\partial\ttime}\,.
\end{equation}

\noindent Again, we cancel the term in square brackets using the continuity
equation. We then integrate equation~(50) over all space. For a system with 
finite extent, the divergence term vanishes. We now introduce the following
definitions for the total kinetic energy
and total thermal energy of the system:
\begin{eqnarray}
\tilde T&\equiv&{1\over2}\int\rho\tvv^2d^3\trr\,,\\
\tilde E&\equiv&\int\trho\tuth d^3\trr\,.
\end{eqnarray}

\noindent With these definitions, equation~(50) reduces to
\begin{equation}
{d\tilde T\over d\ttime}=-{d\tilde E\over d\ttime}-{\cal H}(3\gamma-5)
\tilde E-\int\tphi\,{\partial(\trho-\tbrho)\over\partial\ttime}d^3\trr\,.
\end{equation}

\noindent Notice that integration over all space has transformed the partial
time derivatives into total time derivatives, and that we introduced the
constant mean density $\tbrho$ in the last term. To compute
the last term in equation~(53), we first solve the Poisson equation~(38),
and get
\begin{equation}
\tphi(\tr)=-{3af_n\over2\pi\tbrho}\int{\trho(\tr')-\tbrho
\over|\tr-\tr'|}d^3\trr'\,.
\end{equation}

\noindent We can easily show that this is indeed the correct solution by 
directly comparing equations~(38) and~(54) to their Eulerian counterparts.
With this solution, the last term in equation~(53) becomes
\begin{equation}
-\int\tphi\,{\partial(\trho-\tbrho)\over\partial\ttime}d^3\trr
=-{a\over2}{d\over d\ttime}\biggl[
{\int\tphi(\tr)\big[\trho(\tr)-\tbrho\big]d^3\trr
\over a}\biggr]\,.
\end{equation}

\noindent We then introduce the following definition for the gravitational
potential energy of the system:
\begin{equation}
\tilde U\equiv{1\over2}\int\tphi(\trho-\tbrho)d^3\trr\,.
\end{equation}

\noindent Equation~(53) reduces to
\begin{equation}
{d\over d\ttime}(\tilde T+\tilde E+\tilde U)={\cal H}\Big[
(5-3\gamma)\tilde E+\tilde U\Big]\,.
\end{equation}

\noindent This is the final form of the cosmic energy equation in supercomoving
variables. The corresponding equation in comoving variables (see P80,
eq.~[24.19]) is significantly more complicated.

\section{SOLUTIONS OF THE FRIEDMANN EQUATION}

\subsection{Cases with Elementary Solutions}

The evolution of the cosmological background is described by the Friedmann 
equation~(4), written in supercomoving variables in equation~(32).
For convenience, we rewrite equation~(32) as follows,
\begin{equation}
\biggl({dx\over d\tau}\biggr)^2={4(1-\Omo-\Omx)\over\Omega_0}x^4
+4x^3+{4\Omx\over\Omo}x^{6-n}\,,
\end{equation}

\noindent where
\begin{eqnarray}
x&=&a_0^{-1}a\,,\\
\tau&=&(a_0f_n)^{1/2}\ttime\,.
\end{eqnarray}

\noindent The solution of this equation is
\begin{equation}\tau={1\over2}\int dx\biggl({1-\Omo-\Omx\over\Omo}x^4
+x^3+{\Omx\over\Omo}x^{6-n}\biggr)^{-1/2}\,.
\end{equation}

\noindent This integral has no elementary solution for arbitrary values
of the parameters~$\Omo$, $\Omx$, and~$n$. There are, however,
several physically interesting cases for which elementary solutions exists.
We can identify the cases that have or might have an elementary solution by 
studying the properties of equation~(61). 

First, all known 
cases which are physically relevant have $0\leq n\leq 4$, with $n$ integer.
Hence, we can rewrite equation~(61) as
\begin{equation}
\tau={1\over2}\int {dx\over x}P(x)^{-1/2}\,,
\end{equation}

\noindent where
\begin{equation}
P(x)={1-\Omo-\Omx\over\Omo}x^2+x+{\Omx\over\Omo}x^{4-n}
\end{equation}

\noindent is a polynomial of degree~4 or less. If $n=2$, 3, or~4,
$P(x)$~is a second degree polynomial, and equation~(62) always has
an elementary solution. If $n=1$ (domain walls) or~$n=0$ 
(cosmological constant), then $P(x)$ is a polynomial of degree~3 or~4,
the integral in equation~(62) is elliptic, and does not have an elementary
solution in general. There are special cases, however. Suppose that $P(x)$
has a double root at some particular value~$x=x_{\rm st}$. We can then 
rewrite~$P(x)$ as
\begin{equation}
P(x)=(x-x_{\rm st})^2Q(x)\,,
\end{equation}

\noindent where $Q(x)$ is a linear polynomial if~$n=1$
or a quadratic polynomial if~$n=0$. 
Substituting equation~(64) into equation~(62), we get
\begin{equation}
\tau={1\over2}\int {dx\over x|x-x_{\rm st}|}Q(x)^{-1/2}\,,
\end{equation}

\noindent for which there is always an elementary solution. The 
necessary condition for the polynomial $P(x)$ to have
a double root is that $P(x)$ and its first derivative
both vanish at $x=x_{\rm st}$, which becomes
\begin{eqnarray}
&&\biggl({1-\Omo-\Omx\over\Omo}\biggr)x_{\rm st}^2
+x_{\rm st}+{\Omx\over\Omo}x_{\rm st}^{4-n}=0\,,\\
&&2\biggl({1-\Omo-\Omx\over\Omo}\biggr)x_{\rm st}
+1+(4-n){\Omx\over\Omo}x_{\rm st}^{3-n}=0\,.
\end{eqnarray}

\noindent We can eliminate $x_{\rm st}$ from these equations to get
a condition relating $\Omo$ and~$\Omx$. We get,
after some algebra,
\begin{equation}
1-\Omo-\Omx=\cases{
-3\biggl(\displaystyle{\Omx\Omo^2\over4}\biggr)^{1/3}\,,&$n=0\,;$\cr
-2(\Omx\Omo)^{1/2}\,,&$n=1\,.$\cr}
\end{equation}

\noindent The value of $x_{\rm st}$ is then given by
\begin{equation}
x_{\rm st}=\cases{
\biggl(\displaystyle{\Omo\over2\Omx}\biggr)^{1/3}\,,&$n=0\,;$\cr
\biggl(\displaystyle
{\Omo\over\Omx}\biggr)^{1/2}\,,&$n=1\,.$\cr}
\end{equation}

\noindent The physical meaning of $x_{\rm st}$ is easily understood.
By combining equations~(58), (66), and (67), we can show that
$\dot a$ and $\ddot a$ both vanish at $x=x_{\rm st}$;
hence the universe is {\it static} at $a=a_0x_{\rm st}$.
The case $n=0$ is known as the Einstein static model. In this model,
the repulsive force caused by the cosmological constant 
exactly cancels the gravitational force. As we see, the possibility
of a repulsive force canceling gravity also exists for domain walls
(this is also shown by eq.~[14], in which the X~component 
acts as a negative density for $n<2$ models). 
Since the universe is clearly expanding at present, $x$ must
be smaller than the equilibrium value~$x_{\rm st}$. Equations~(58)
and~(68) then describe a universe which decelerates and asymptotically
approaches a static universe as $x$ gets closer to $x_{\rm st}$.

\subsection{Families of Solutions}

Equation (61) (or~[32]) relates the cosmic scale factor $a$ to our 
supercomoving time variable $\ttime$ in any particular model. According to
this equation, models with different values of $\Omo$ and $\Omx$ can often
be grouped into ``families'' within which the dependence of $a$ on $\ttime$
is the same, as follows.

There is a basic difference which occurs between those cases for which
$P(x)$ defined by equation~(63) reduces to only two terms and those cases
for which $P(x)$ has three terms. Let us call the former case, category~1,
while the latter is category~2. For cases in category~1, it is always possible
to find solutions with no explicit dependences upon the cosmological
parameters $\Omo$ and $\Omx$.
If $P(x)$ contains only two terms,
the change of variable $x=by$, with an appropriate choice for~$b$,
will make the coefficients of these two terms identical. We can then
move that coefficient to the left hand side of equation~(62) and eliminate
it by rescaling~$\tau$. Cases for which $P(x)$ has only two terms
include matter-dominated models ($\Omx=0$, as described in S80), any
zero-curvature model ($1-\Omo-\Omx=0$) regardless of the presence of
an additional uniform background component, models with cosmic strings ($n=2$),
and models with a nonrelativistic uniform component ($n=3$). 
In each of these cases in category~1, the solution for $a(\ttime\,)$ as
a function of $\ttime$ reduces to a finite number of functions which describe
the infinite number of combinations of values for $\Omo$ and $\Omx$ which can
occur in each case. We present these solutions in \S7.3. In all other 
cases, $P(x)$ contains 3~terms (category~2), and 
there are an infinite number of solutions for the dependence of $a(\ttime\,)$
on $\ttime$, but we can still identify combinations of $\Omo$ and $\Omx$
which share a common solution. To find these combinations, we set
$x=(\Omo/\Omx)^{1/(3-n)}y$ in equation~(63), which
makes the coefficients of the last two terms identical. We get
\begin{equation}
P(y)=\biggl({\Omo\over\Omx}\biggr)^{1/(3-n)}\biggl[
{(1-\Omo-\Omx)\Omo^{(n-2)/(3-n)}\over\Omx^{1/(3-n)}}y^2
+y+y^{4-n}\biggr]\,.
\end{equation}

\noindent As long as the following condition is met:
\begin{equation}
\kappa\equiv{(1-\Omo-\Omx)\Omo^{(n-2)/(3-n)}\over\Omx^{1/(3-n)}}={\rm const}
\,,
\end{equation}

\noindent the solution will not depend upon $\Omo$ and $\Omx$,
For any combination of $\Omo$ and $\Omx$ which satisfies equation~(71), there
is a single dependence of $a(\ttime\,)$ on $\ttime$, which is different
for different values of $\kappa$. There are, therefore, an infinite
number of such cases, corresponding to the infinite range of values of 
$\kappa$. The particular cases with $n=0$ or 1 for which closed-form, 
analytical solutions exist (eq.~[68]) do satisfy the condition in 
equation~(71).

Henceforth, we shall refer to the combination of values of $\Omo$, $\Omx$,
and $n$ which share a common solution for the dependence of $a(\ttime\,)$
on $\ttime$ as ``families.'' The matter-dominated cases ($\Omx=0$) considered
by Shandarin (S80) belong to category~1 of the cases described above.
As shown by Shandarin, these models which are
parametrized by $\Omo$ are, in supercomoving variables,
fully described by three families, closed,
flat, and open, and within each family, the dependence of $a(\ttime\,)$ on
$\ttime$ is identical, independent of the value of $\Omo$. We have shown above
that this concept of ``families'' can be extended to include the much
wider range of cosmological models considered here, parametrized by
$\Omo$, $\Omx$, and $n$. For category~1 models, of which the matter-dominated
are one example, there are also just a finite number of families per model
(i.e. for each value of $n$) as in the matter-dominated case. For category~2
models, however, each value of $\kappa$ corresponds to a different family.
This might look like a mathematical trick at first sight,
where we use equation~(71) to group different models that appear to be
unrelated. However, the fact that, within a given family, the models
will be identical in supercomoving variables indicates that the various models
assembled into the same family are indeed deeply related. For instance, for any
particular value of~$n$, the flat (zero-curvature) models constitute a family,
defined by~$\kappa=0$. Hence, all flat models with the same value of~$n$ 
will have the same solution for $a(\ttime\,)$ in supercomoving variables. 
Other physically interesting families include
critical models with a cosmological constant, defined by~$n=0$,
$\kappa=-3/4^{1/3}$,
and critical models with domain walls, defined by~$n=1$, $\kappa=-2$.
Now that we have derived equation~(71) and defined the concept of family,
we are ready to solve the Friedmann equation for the various cases
which have elementary solutions.

\subsection{Solutions}

\subsubsection{Matter-Dominated Universe $(\Omx=0)$}

These are the only models
considered in S80. The solutions of the Friedmann equation are
\begin{equation}
a=\cases{
(\ttime\,^2-1)^{-1}\,,&$\Omo<1\,;$\cr
\ttime^{-2}\,,&$\Omo=1\,;$\cr
(\ttime\,^2+1)^{-1}\,,&$\Omo>1\,,$\cr}
\end{equation}

\noindent where the
present value~$a_0$ of the scale factor is
\begin{equation}
a_0=\cases{
(1-\Omo)/\Omo\,,&$\Omo<1\,;$\cr
1\,,&$\Omo=1\,;$\cr
(\Omo-1)/\Omo\,,&$\Omo>1\,.$\cr}
\end{equation}

\noindent
These values of $a_0$ were chosen in order to eliminate the
dependences upon $\Omo$ in equation~(72).
These supercomoving-variable
solutions are remarkably simple compared with their comoving-variable
counterparts. 
In equation~(72), $\ttime$ is negative and increases with time from
the value~$\ttime=-\infty$ at the Big Bang. For the flat ($\Omo=1$)
model, $\ttime=-1$ at the present, and approaches~0 as the universe expands
to infinite radius.  
For the open ($\Omo<1$) model, $\ttime<-1$ at the present,
and approaches~$-1$
as the universe expands to infinite radius. For the closed ($\Omo>1$) model,
$\ttime<0$ at present, the universe reaches a maximum expansion
at $\ttime=0$ and recollapses at $\ttime=\infty$
(``the Big Crunch''). The solutions of equation~(72) for the
cosmic scale factor $a(\ttime\,)$ as a function of $\ttime$ are plotted
in Figure~1a (solid curves). The solid dot on the curve for the
flat model indicates the present. For all other models, the location of the 
present is a function of~$\Omo$. In Figure~1b, we 
have plotted the ``supermoving 
Hubble parameter'' ${\cal H}=a^{-1}da/d\ttime$ defined by equation~(31),
also using solid curves.

\subsubsection{Universe with Infinite Strings $(n=2)$}

This case is extremely simple. Setting $n=2$ in equation~(61) has exactly the 
same effect as setting $\Omx=0$. Hence, the solutions for a 
universe with strings are exactly the same as 
equations~(72) and~(73) above for a matter-dominated
universe, independent of the value
of $\Omx$. This results from the fact that a uniform background term
with $n=2$ behaves exactly like the curvature term in the Friedmann
equation. Hence, the solid curves plotted in Figure~1 for the 
matter-dominated models apply to these models as well.

\subsubsection{Universe with a Nonclumping, Nonrelativistic Component $(n=3)$}

This case is also very simple. We define an effective density parameter
$\Omo'$ as
\begin{equation}
\Omo'=\Omo+\Omx\,.
\end{equation}

\noindent Equation~(61) then reduces to
\begin{equation}
\tau={1\over2}\biggl({\Omo\over\Omo'}\biggr)^{1/2}
\int dx\biggl({1-\Omo'\over\Omo'}x^4+x^3\biggr)^{-1/2}\,.
\end{equation}

\noindent This equation has exactly the same form as in the case of a
matter-dominated universe, except for the extra 
factor~$(\Omo/\Omo')^{1/2}$ in front of the integral. 
However, this extra factor disappears after we eliminate $\tau$ using
equations~(30) and~(60), and we get exactly the same solutions as for the
matter dominated case,
\begin{equation}
a=\cases{(\ttime\,^2-1)^{-1}\,,&$\Omo+\Omx<1\,;$\cr
\ttime^{-2}\,,&$\Omo+\Omx=1\,;$\cr
(\ttime\,^2+1)^{-1}\,,&$\Omo+\Omx>1\,;$\cr}
\end{equation}

\noindent where the
present value of the scale factor is given by
\begin{equation}
a_0=\cases{
\displaystyle{1-\Omo-\Omx\over\Omo+\Omx}\,,&$\Omo+\Omx<1\,;$\cr
1\,,&$\Omo+\Omx=1\,;$\cr
\displaystyle{\Omo+\Omx-1\over\Omo+\Omx}\,,&$\Omo+\Omx>1\,.$\cr
}
\end{equation}

This was the reason for treating the case $n=3$ differently
in equations~(25) and~(60). Otherwise, the solutions would have shown
dependences upon $\Omo$ and $\Omx$. The solutions~(76)
and~(77) are identical to the solutions~(72) and~(73) for the matter-dominated
cases (and hence, the solid curves plotted in Figure~1 apply to
these models as well) if we replace $\Omo$
in the latter by $\Omo'$.
This makes sense since, for $n=3$, the clumping,
nonrelativistic and nonclumping, nonrelativistic X~components
are indistinguishable in the Friedmann equation. They are distinguishable
in the Poisson equation, however, since one component can contribute
to the peculiar gravitational potential while the other cannot.
Hence, we end up with a different form for the Poisson equation,
which is the price we must pay for having no 
dependence in equation~(76)
upon the cosmological parameters. 

This modification
of the Poisson equation is actually physically interesting. We can rewrite
equation~(39) as
\begin{equation}
|\tgrad^2\tphi|=
6a\biggl|{\trho+\tbrho_X
\over\tbrho+\tbrho_X}-1\biggr|<
6a\biggl|{\trho\over\tbrho}-1\biggr|\,.
\end{equation}

\noindent Since the solution $a(\ttime\,)$ is the same for both matter 
dominated and $n=3$ models, the interpretation of the inequality in 
equation~(78) is that {\it the presence of the X~component effectively 
weakens the peculiar gravity}.
This explains, for example, why the growth of density fluctuations in
the baryons and CDM is reduced in a universe which includes massive
neutrinos, the Cold + Hot Dark Matter (CHDM) model, compared with the
growth in a CDM model with the same $\Omo$, for wavelengths smaller than
the neutrino free-streaming length. In the CHDM model, the
growth of small-wavelength fluctuations occurs as if the neutrinos 
corresponded to a nonclumping, nonrelativistic background with
$\Omega_\nu=\Omx$, until very late times when long wavelength neutrino
density fluctuations achieve significant growth too.

\subsubsection{Universe with a Radiation Background $(n=4)$}

The solution of equation~(61) for $n=4$ is
\begin{equation}
a=e^{-2\ttime}\Biggl[{(e^{-2\ttime}-1)^2\over4}
-{\Omx(1-\Omo-\Omx)\over\Omo^2}\Biggr]^{-1}\,,
\end{equation}

\noindent where the
present value~$a_0$ of the scale factor is
\begin{equation}
a_0={\Omo\over\Omx}\,.
\end{equation}

\noindent As in other cases,
this solution gives $a=0$ (Big Bang) for $\ttime=-\infty$. 
The behavior of 
the solution depends upon the curvature of the universe. 
If~$\Omo+\Omx<1$, $a$ becomes infinite when the
square bracket in equation~(79) is zero. This gives a maximum
value for~$\ttime$,
\begin{equation}
\ttime(a=\infty)=-{1\over2}\ln\Biggl\{1+
\biggl[{4\Omx(1-\Omo-\Omx)\over\Omo^2}\biggr]^{1/2}
\Biggr\}\,.
\end{equation}

If~$\Omega_0+\Omx>1$, the square bracket in equation~(79)
can never be~0, implying that $a$ can never be infinite. 
This case corresponds to that of a bound universe. The Big Crunch ($a=0$)
occurs at $\ttime=+\infty$. The time at turnaround can be found by setting
$da/d\ttime=0$. We find
\begin{equation}
\ttime_{\rm turnaround}=-{1\over4}\ln\Biggl[1+{4\Omx(\Omo+\Omx-1)\over\Omo^2}
\Biggr]\,.
\end{equation}

Unlike in the previous cases, we have not succeeded in obtaining a solution
that has no dependences upon the parameters~$\Omo$ and~$\Omx$. However, 
the solution~(79) will be the same for different models with the same value
of~$\kappa=\Omx(1-\Omo-\Omx)/\Omo^2$, which is precisely what
equation~(71) predicted. The essential difference between this case and the
previous cases is that matter-dominated models,
models with cosmic strings, and models with a uniform nonrelativistic 
component are one-parameter models (the parameter being~$\Omo$ for the 
first two cases, and~$\Omo'\equiv\Omo+\Omx$ for the latter case). 
It is then possible, 
as we saw, to find solutions with no explicit dependence upon that parameter.
The radiation model, however, is a two-parameter model, and we cannot 
eliminate these parameters from the solution. The solution to this problem,
as we saw in \S7.2,
is to consider models such as the radiation model, not as two-parameter
models, but as an ensemble of one-parameter models, each model being 
characterized by the value of the constant~$\kappa$ in equation~(71). 
Equation~(79) then reduces to
\begin{equation}
a=e^{-2\ttime}\Biggl[{(e^{-2\ttime}-1)^2\over4}
-\kappa\Biggr]^{-1}\,.
\end{equation}

\noindent For any particular value of $\kappa$, we are in a family
for which the cosmological parameters~$\Omo$ and~$\Omx$ are
related by equation~(71). All these models have the same solution,
equation~(83), independent of the value of the cosmological parameters.
We have plotted the solution for flat models ($\kappa=0$) in Figure~1
(dotted curves). The radiation component dominates
at early times, which explains why this model has a different behavior
at early time compared to all other models.
The most interesting family is the one for which the spatial 
curvature is zero, which corresponds to $\kappa=0$.
In the limit~$\ttime\rightarrow0^-$, we recover the solution
$a=\ttime^{-2}$, for the flat matter-dominated model, since the relative
contribution of the radiation component decays as the universe expands.

\subsubsection{Critical Universe with a Nonzero Cosmological Constant 
$(n=0$, $\Omx=\Omega_{X0,\rm crit})$}

A universe with a positive cosmological constant is marginally bound if
$\Omo$ and~$\Omx$ satisfy equation~(68) for~$n=0$.
We substitute this expression into equation~(61). The resulting integral
has an elementary solution, which is
\begin{equation}
\ttime=-\biggl(1+{1\over a}\biggr)^{1/2}-{1\over3^{1/2}}\ln
{a+1-\bigl[3a(a+1)\bigr]^{1/2}\over
 a+1+\bigl[3a(a+1)\bigr]^{1/2}}\,,
\end{equation}

\noindent where the present scale factor is given by
\begin{equation}
a_0=\biggl({\Omx\over4\Omo}\biggr)^{1/3}\,.
\end{equation}

\noindent Unfortunately, we cannot invert equation~(84) to express
$a$ as a function of $\ttime$ (except numerically). In this model,
$\ttime=-\infty$ at the Big Bang, and goes to $\ttime=+\infty$ as
the scale factor~$a$ approaches its maximum value $a_{\max}=1/2$.
We have plotted this solution in Figure~1.

\subsubsection{Critical Universe with Domain Walls 
$(n=1$, $\Omx=\Omega_{X0,\rm crit})$}

A universe with domain walls is marginally bound if
$\Omo$ and~$\Omx$ satisfy equation~(68) for~$n=1$.
We substitute this expression in equation~(61). The resulting integral
has an elementary solution, which is
\begin{equation}
\ttime=-{1\over a^{1/2}}+\tanh^{-1}a^{1/2}\,,
\end{equation}

\noindent where the present scale factor is given by
\begin{equation}
a_0=\biggl({\Omx\over\Omo}\biggr)^{1/2}\,.
\end{equation}

\noindent Again, we cannot invert equation~(86) to express
$a$ as a function of $\ttime$. In this model,
$\ttime=-\infty$ at the Big Bang, and goes to $\ttime=+\infty$ as
the scale factor~$a$ approaches its maximum value $a_{\max}=1$.
We have plotted this solution in Figure~1

\subsubsection{Flat Models with Domain Walls or a Cosmological Constant
$(\Omo+\Omx=1$, $n=1$ {\rm or} $0)$}

These cases are interesting and important for several reasons. First,
the flatness of the cosmological model is usually regarded as a
requirement of the standard inflationary scenario. 
Second, as summarized by Ostriker \&~Steinhardt (1995), the flat
model with nonzero cosmological constant and Cold Dark Matter is
of great interest as an explanation for several observations which
are otherwise discordant with the standard flat CDM model
without a cosmological constant. For example, the presence of 
a cosmological constant (or domain walls)
increases the age of the universe for a given observed~$H_0$,
helping to solve the so-called age problem.
As a practical matter, therefore, it is convenient that, 
according to the discussion
presented in \S7.2, solutions of the Friedmann equation
for flat models, elementary or not, will have no explicit dependence
upon the cosmological parameters.

Equation~(61) for flat models reduces to
\begin{equation}
\tau={1\over2}\int dx\biggl(x^3+{\Omx\over\Omo}x^{6-n}\biggr)^{-1/2}\,.
\end{equation}

\noindent Using equations~(59) and~(60), we can write
this integral as
\begin{equation}
\ttime={1\over2}\int_1^a {dy\over y^{3/2}(1+y^{3-n})^{1/2}}\,.
\end{equation}

\noindent where the present scale factor is defined by
\begin{equation}
a_0=\biggl({\Omx\over\Omo}\biggr)^{1/(3-n)}\,.
\end{equation}

\noindent By writing equation~(89) as a definite integral, we have fixed
the value of the integration constant by imposing (arbitrarily) 
$a=1$ at $\ttime=0$. As expected, equation~(89) has no 
explicit dependence upon~$\Omo$
and~$\Omx$. Unfortunately, this integral has no elementary solutions for
the cases~$n=0$ and~$n=1$, and thus has to be evaluated numerically. 
In the limit~$a\ll1$, the term $y^{3-n}$ can be neglected, and we recover the
solution~$\ttime=-a^{1/2}$, $a=\ttime^{-2}$ of the matter-dominated flat 
universe (eq.~[72]). This was expected, since the contribution of the
uniform component is negligible at early times for $n<3$. The value 
of~$\ttime$ at present is obtained by integrating equation~(89) numerically
for $a=a_0$. The results are given in Table~1 for
various cases. Finally, $\ttime$~approaches a finite value as the universe
expands to infinite radius. This value can be computed numerically by
integrating equation~(89) with~$a=\infty$. The results are also given in 
Table~1.

\subsubsection{Solution Families}

The results presented in this section are summarized in Figure~2,
where we plot the relationships between $\Omo$ and $\Omx$ which
define family membership for each value of $n$,
for models with $n=0$, 1, 2, 3, and~4.
For models with $n=0$, 1, and~4, families are defined by equation~(71), with 
each possible value of $\kappa$ corresponding to a particular family. 
Families are thus represented by curves of constant $\kappa$. We 
have plotted
several of these curves in Figure~2a, 2b, 2c, and 2f, and indicated the
corresponding values of $\kappa$. Thick dashed curves indicate particular 
families that were discussed in this section
(flat and critical models with a cosmological constant or domain walls,
and flat models with radiation). 
The cases $n=2$ and $n=3$ are different. As we saw, there are only 3 families
for these models. These families are represented in Figure~2d and~2e by
the dashed line, and the half-planes on either one side of it.
The dashed areas in Figure 2a and 2c represent
regions in parameter space that are excluded by the existence of the
Big Bang. In these regions, the Friedmann equation~(4) predicts a
minimum value $a=a_{\min}>0$ for the scale factor. For $a<a_{\min}$,
equation~(4) gives $(\dot a)^2<0$.  

The solutions presented in this section describe the evolution of the
cosmological background. The next step consists of describing the growth
of fluctuations leading to the formation of large-scale structure. This
problem is much too complex to be solved analytically for the general
case. In the next two sections, we consider two particular cases of
great cosmological importance for which analytical solutions exists,
either because various terms in the fluid equations can be neglected,
or because the problem has a high degree of symmetry. These two
problems are: (1) the growth of fluctuations in the linear regime, and
(2) the growth of a 1D, plane-wave density fluctuation from the linear
to the nonlinear regime, leading to the formation of a collisionless
pancake. The extension of these results to the nonlinear regime for a
general distribution of initial density fluctuations is then made possible
by the well-known Zel'dovich approximation presented in supercomoving 
variables is \S10.

\section{LINEAR PERTURBATION THEORY}

\medskip

We can use the fluid conservations equations in supercomoving 
variables to describe the growth of small-amplitude density
fluctuations in the early universe. 
First, we decompose the density~$\trho$ and the pressure~$\tp$
into a uniform and a space-varying component,
\begin{eqnarray}
\trho&=&\tbrho(1+\delta)\,,\\
\tp&=&\tilde{\bar p}+\tilde{\delta p}\,,
\end{eqnarray}

\noindent where $\tbrho$ and $\tilde{\bar p}$ are the mean
values of the density and pressure. In supercomoving
variables, $\tbrho$ is constant and
$\tilde{\bar p}$ is a function of time only (see eq.~[44])
(and $\tilde{\bar p}=\hbox{constant}$ if $\gamma=5/3$). 
We substitute equations~(91) and~(92) into the
fluid equations. These equations reduce to
\begin{eqnarray}
&&{\partial\delta\over\partial\ttime} 
+\tgrad\cdot\big[\tv(1+\delta)\bigr]=0\,,\\
&&{\partial\tv\over\partial\ttime} +\tv\cdot\grad\tv=
-{\tgrad\tilde{\delta p}\over\tbrho(1+\delta)}-\tgrad\tphi\,,\\
&&\tgrad^2\tphi=6af_n\delta\,.
\end{eqnarray}

\noindent A trivial case which satisfies these
equations is $\delta=0$, $\tv=0$,
$\tphi=0$, $\tilde{\delta p}=0$, which corresponds to a universe 
with no fluctuation.
Assuming that the fluctuations are small, we can solve equations~(93)--(95),
in general,
using perturbation theory. After dropping second order terms, equations~(93)
and~(94) reduce to
\begin{eqnarray}
&&{\partial\delta\over\partial\ttime} +\tgrad\cdot\tv=0\,,\\
&&{\partial\tv\over\partial\ttime} =-{\tgrad\tilde{\delta p}\over\tbrho}
-\tgrad\tphi\,.
\end{eqnarray}

\noindent (Eq.~[95] is already linear.)
To solve these equations, we take the divergence of equation~(97),
then use equation~(96) to eliminate~$\tv$, and equation~(95)
to eliminate~$\tphi$. We get
\begin{equation}
{\partial^2\delta\over\partial\ttime^2}={\tgrad^2\tilde{\delta p}\over\tbrho}
+6af_n\delta\,.
\end{equation}

\noindent To solve this equation, we need a relationship between the
pressure fluctuation~$\tilde{\delta p}$ 
and density contrast~$\delta$. For adiabatic
perturbations, 
$\tgrad^2\tilde{\delta p}=\tilde{\bar c}_s^2\tbrho\tgrad^2\delta$ to first
order, where~$\tilde{\bar c}_s=(\gamma\tilde{\bar p}/\tbrho)^{-1/2}$
is the mean sound speed. Equation~(98) reduces to
\begin{equation}
{\partial^2\delta\over\partial\ttime^2}=\tilde{\bar c}_s^2\tgrad^2\delta
+6af_n\delta\,.
\end{equation}

\subsection{Critical Jeans Length}

The density contrast $\delta$ can be expressed as a sum of plane waves,
\begin{equation}
\delta=\sum_\kbf\delta_\kbf e^{i\kbf\cdot\tr}\,.
\end{equation}

\noindent
We substitute equation~(100) in equation~(99). The terms for each value 
of~$\kbf$ must cancel out separately. We then get a separate equation for
each mode~$\kbf$,
\begin{equation}
{\partial^2\delta_\kbf\over\partial\ttime^2}=
6af_n\Biggl[1-\biggl({\tilde\lambda_J\over\tilde\lambda}
\biggr)^2\Biggr]\delta_\kbf\,,
\end{equation}

\noindent where $\tilde\lambda=2\pi/|\kbf|$ is the wavelength of the mode, and
$\tilde\lambda_J$ is the Jeans wavelength, defined by
\begin{equation}
\tilde\lambda_J=2\pi(6af_n)^{-1/2}\tilde{\bar c}_s\,.
\end{equation}

\noindent Perturbations can grow only if 
$\tilde\lambda>\tilde\lambda_J$. Modes with
$\tilde\lambda<\tilde\lambda_J$ result in sound waves. The mean sound speed
is constant for $\gamma=5/3$ models, and a function of $a$ alone for
$\gamma\neq5/3$ models. Thus, the only time dependence
in equation~(102) is contained in the scale factor~$a(\ttime\,)$. This implies
that within any given family, the value of the Jeans length is independent
of the cosmological parameters. For instance, we can compute the Jeans length
for matter-dominated models by substituting equation~(72) into equation~(102).
We get
\begin{equation}
\tilde\lambda_J={2\pi\over6^{1/2}}\tilde{\bar c}_{s,0}a_0^{(3\gamma-5)/2}
\times\cases{
(\ttime\,^2-1)^{3\gamma/2-2}\,,&$\Omo<1\,;$\cr
\ttime\,^{3\gamma-4}\,,&$\Omo=1\,;$\cr
(\ttime\,^2+1)^{3\gamma/2-2}\,,&$\Omo>1\,.$\cr}
\end{equation}

We note that, if the temperature of the baryon-electron gas 
remains coupled to that of the cosmic microwave radiation background by
Compton scattering, then $\gamma=4/3$ is the appropriate value to
use in equation~(43) and $\tilde{\bar c}_s\propto a^{1/2}$. In that case,
equation~(102) indicates that $\tilde\lambda_J$ is independent of time.
This means that the baryon Jeans {\it mass} $M_J$ (in physical units, where
$M_J\propto\bar\rho\lambda_J^3$) must be constant in time, as well,
in that case. This latter result is a well-known one for the 
Einstein-de~Sitter model, and is relevant to the postrecombination epoch 
between $z\sim10^3$ and $z\sim10^2$ for mean baryon densities consistent with
Big Bang nucleosynthesis constraints for the observed values of the Hubble
constant (see, for example, Shapiro, Giroux, \& Babul 1994). 
However, by solving the problem in supercomoving variables, we have
generalized this result to a very much wider class of models.

\subsection{Zero Pressure Solutions}

In the limit~$\lambda\gg\lambda_J$, equation~(99) reduces to
\begin{equation}
{\partial^2\delta\over\partial\ttime^2}=6af_n\delta\,.
\end{equation}

\noindent This equation has a slightly simpler form in supercomoving 
variables than in comoving variables. However, both equations are 
equally complicated
to solve. Indeed, the first step in solving equation~(104) is to replace the
independent variable~$\ttime$ by~$x=a/a_0$ using equations~(25), (32),
and~(33). Equation~(104) reduces to
\begin{eqnarray}
\bigl[(1-\Omo-\Omx)x^3&+&\Omo x^2+\Omx x^{5-n}\bigr]\delta''\nonumber\\
&+&\Biggl[2(1-\Omo-\Omx)x^2+{3\over2}\Omo x+\biggl(3-{n\over2}\biggr)
\Omx x^{4-n}\Biggr]\delta'={3\Omo\over2}\delta\,,
\end{eqnarray}

\noindent where a prime stands for~$d/dx$ (notice that $f_n$ has 
disappeared from the equation). 
This equation is the same in supercomoving and comoving variables. It has two
independent solutions, a decaying solution~$\delta_-$ and a growing 
solution~$\delta_+$. We shall focus on the growing solution, which is
responsible for the formation of structures.
Numerous solutions of equation~(105) have been published. Hence,
we only need to reexpress these known solutions in terms of supercomoving 
variables, by eliminating the scale factor~$a(\ttime\,)$ using the
solutions derived in \S7.3. 

\subsubsection{Matter-Dominated Universe $(\Omega_{X0}=0)$}

The solutions for matter-dominated models
are given in Weinberg (1972) and Groth \&~Peebles (1975). 
In terms of supercomoving variables these solutions are
\begin{equation}
\delta_+=\cases{
\displaystyle{5\over2}\Biggl[1+3(\ttime\,^2-1)
\biggl(1-\ttime\tanh^{-1}{1\over\ttime}\biggr)\Biggr]\,,&$\Omo<1$;\cr
\ttime^{-2}\,,&$\Omo=1$;\cr
\displaystyle{5\over2}\Biggl[-1+3(\ttime\,^2+1)
\biggl(1-\ttime\tan^{-1}{1\over\ttime}\biggr)\Biggr]\,,&$\Omo>1$;\cr}
\end{equation}

\noindent where the factors of $5/2$ were chosen such
that all solutions reduce to~$\delta_+=\ttime^{-2}$ at
early time. These solutions were 
given in S80
(Actually, S80 eq.~[15] contains some typographical errors. We have
corrected this in our eq.~[106] above). 

\subsubsection{Universe with Infinite Strings $(n=2)$}

As in the case of the Friedmann equation, setting~$n=2$ in
equation~(105) has the same effect as setting~$\Omx=0$, hence the
solutions~(106) also apply to models with infinite strings.

\subsubsection{Universe with a Nonzero Cosmological Constant $(n=0)$}

The general solutions of the perturbation equation for models
with a nonzero cosmological constant are given in Edwards \&~Heath (1976),
P80, Martel (1991), Lahav et al (1991), and Bildhauer, Buchert, \&~Kasai 
(1992) (see also Carroll, Press, \&~Turner 1992)
For flat models, the solution is
\begin{equation}
\delta_+={5\over2}\biggl(1+{1\over a^3}\biggr)^{1/2}\int\limits_0^a{w^{3/2}dw
\over(1+w^3)^{3/2}}\,,
\end{equation}

\noindent
and for critical models,
\begin{equation}
\delta_+=5
\biggl({1-3a+4a^3\over4a^3}\biggr)^{1/2}\int\limits_0^a{w^{3/2}dw
\over(1+w)^{3/2}(1-2w)^3}\,.
\end{equation}

\noindent 
Since $a$ is not an elementary function of~$\ttime$ for these models,
we cannot express $\delta_+$ as a function of~$\ttime$
in closed form. Notice, however, that
supercomoving variables still played an important role. 
If we had not been using 
the definitions of $a_0$ given by equations~(85) and~(90), the 
solutions~(107) and~(108) would show explicit dependences 
upon~$\Omo$ and~$\Omx$, as they normally do in comoving variables.

\subsubsection{Universe with a Radiation Background $(n=4)$}

Solutions for models with a radiation background were derived by M\'esz\'aros
(1974) and Groth \&~Peebles (1975). In supercomoving variables, 
the growing solution
for flat models (that is, $\kappa=0$ in eq.~[83]) is
\begin{equation}
\delta_+=1+{6e^{-2\ttime}\over(e^{-2\ttime}-1)^2}\,.
\end{equation}

\subsubsection{Universe with a Nonclumping, Nonrelativistic Component $(n=3)$}

Models with a nonclumping, nonrelativistic component ($n=3$) were discussed
by Bond, Efstathiou, \&~Silk (1980) and Wasserman~(1981). 
Such models are of interest in the context of models like the Cold+Hot
Dark Matter model (CHDM) in which one component of matter, the
Hot Dark Matter, has a much larger effective Jeans length than the others. 
In that case, while fluctuations are able to grow in the other components
on scales smaller than this effective Jeans length in the HDM, they
do not grow in the HDM. The $n=3$ model, therefore,
becomes a useful approximation for small wavelengths in the CHDM model.
For flat models, the solution is
\begin{equation}
\delta_+=a^m=(-\ttime\,)^{-m/2}\,,
\end{equation}

\noindent where
\begin{equation}
m={1\over4}\Bigl[-1+(1+24\Omo)^{1/2}\Bigr]\,.
\end{equation}

\noindent
For non-flat models, the solutions are not power laws. In the
limit~$a/a_0\ll1$, however, these solutions approach a power law, but
with a different exponent,
\begin{equation}
m={1\over4}\Biggl[-1+\biggl(1+{24\Omo\over\Omo+\Omx}\biggr)^{1/2}\Biggr]\,,
\qquad a\ll a_0\,.
\end{equation}

\noindent Notice that the solutions~(110)--(112) do not approach 
$\delta_+\propto a$
at early time as they do in all models with~$n<3$. 
Hence, there is no preferred normalization for these solutions.

Unlike in the previous cases, the solutions~(110)--(112) 
explicitly depend
upon the density parameter~$\Omo$. In all the
previous cases, the solutions were not power laws. Hence, there was a natural
time scale in each solution, enabling us to rescale the time variable in order
to make these solutions identical. But in the case of equation~(110), 
the solution
is a power law, and therefore has no natural time scale that could be used for
rescaling. Hence, the solutions~(110) for different
values of~$\Omo$ are genuinely
different, and cannot be transformed into one another by a rescaling of the
variables.

\subsubsection{Universe with Domain Walls $(n=1)$}

For models with domain walls ($n=1$), we could not find any published
solution in the literature. We shall therefore derive our own solutions
of equation~(105) for~$n=1$. As in \S7.3, we consider two specific cases,
flat models and critical models.

First, consider as a trial solution
the decaying solution for the Einstein-de~Sitter model
($\Omo=1$, $\Omx=0$), that is, $\delta_-=x^{-3/2}$. We substitute this 
solution into equation~(105), and get
\begin{equation}
{3\over4}(1-\Omo-\Omx)x^{1/2}+{3\over4}\Omx(n-1)x^{5/2-n}=0\,.
\end{equation}

\noindent This trial solution is the correct solution if both terms in 
equation~(113) vanish. The
first term vanishes for flat models. The second term vanishes either
for~$\Omx=0$ or~$n=1$. The former case corresponds to the Einstein-de~Sitter
model, the latter to a model with domain walls. Hence {\it the decaying
mode is the same in a flat universe with domain walls as in a
flat universe with no X~component}. Of course, this is true
only if we express~$\delta_-$ as a function of~$a$ (or~$x\equiv a/a_0$).
The relationship between~$a$ and~$\ttime$ is model-dependent, and so is
the relationship between~$\delta$ and~$\ttime$.

We thus have found the decaying mode for flat models with domain walls. 
To find the growing mode, we set~$\delta=\delta_-\int f(x)dx$ and substitute 
in equation~(105), with~$1-\Omo-\Omx=0$ and $n=1$. We get
a first order equation for~$f(x)$, which is easily solved. The final solution 
is
\begin{equation}
\delta_+={5\over2}
a^{-3/2}\int\limits_0^a{w^{3/2}dw\over(1+w^2)^{1/2}}\,.
\end{equation}

For critical models with domain walls, we could not find any elementary   
solution for equation~(106). However, we can express the growing solution as
an infinite series. Since the solution must take the form~$\delta_+\propto a$
in the limit~$a\ll1$ (this is true for all models with $n<3$), we define
\begin{equation}\delta_+=\sum_{l=0}^\infty c_la^{l+1}\,.
\end{equation}

\noindent
We substitute this solution into the perturbation equation, and solve for the
coefficients. The solution is
\begin{equation}
c_0=1\,,\qquad c_1={8\over7}\,,\qquad
c_l={4l(l+1)c_{l-1}+(-2l^2+l+1)c_{l-2}\over l(2l+5)}\,,\quad l\geq2\,.
\end{equation}

\noindent As in all the previous cases with~$n<3$, we normalize these 
solutions by imposing~$\delta_+=\ttime^{-2}$ at early time.

\section{THE COLLISIONLESS PANCAKE PROBLEM}

For a 1D, planar density perturbation, it is possible to solve the fully 
nonlinear problem exactly, up to the moment of caustic formation, in the
limit in which pressure can be neglected
(Sunyaev \& Zel'dovich 1972; Doroshkievich, Ryabenkii,
\& Shandarin 1973). 
This is the problem which is
sometimes described as the cosmological ``pancake'' problem 
(Zel'dovich 1970). Writing this solution in supercomoving variables serves 
to demonstrate the universal character of the solution, independent of
cosmological model, as follows. We consider, for simplicity, a 1D, 
primordial plane-wave density fluctuation as the initial condition, which is
a solution of the linear perturbation equations discussed in the previous 
section.

In the collisionless limit ($\tp=0$), the fluid conservation equation in
supercomoving form, for plane-symmetric systems, reduce to
\begin{eqnarray}
&&{\partial\trho\over\partial\ttime}
+{\partial\over\partial\tx}(\trho\tvv)=0\,,\\
&&{\partial\tvv\over\partial\ttime}
+\tvv{\partial\tvv\over\partial\tx}
=-{\partial\tphi\over\partial\tx}\,,\\
&&{\partial^2\tphi\over\partial\tx^2}=6af_n\biggl({\trho\over\tbrho}-1
\biggr)\,.
\end{eqnarray}

We now set up initial conditions at time $\ttime_i$
by displacing each mass element from its unperturbed location $\tq$ by
a small amount $\txi_i$, as follows,

\begin{equation}
\tx_i=\tq-\txi_i\,.
\end{equation}

\noindent
For the particular problem of the collisionless pancake, we choose
initial conditions that are periodic,
\begin{equation}
\txi_i={b_i\over2\pi\tk}\sin2\pi\tk\tq\,,
\end{equation}

\noindent where $\tk$ is the wavenumber of the perturbation. The corresponding
initial density profile is given by
\begin{equation}
\trho_i={\tbrho\over1-b_i\cos2\pi\tk\tq}\simeq
\tbrho(1+b_i\cos2\pi\tk\tq)\,,
\end{equation}

\noindent where the last equality is valid in the limit $b_i\ll1$. The
initial density contrast is therefore given by $\delta_i=b_i\cos2\pi\tk\tq$.
Since the density contrast at early times grows according to the solution
of the linear perturbation equations, the solution at early times must reduce 
to
\begin{equation}
\delta(\ttime\,)
=b_i{\delta_+(\ttime\,)\over\delta_+(\ttime_i)}\cos2\pi\tk\tq\,.
\end{equation}

\noindent
Our goal is
to find a solution of the fluid equations~(117)--(119) which reduces to 
equation~(122) at early times. 
The solution is given by
\begin{equation}
\txi(\tx,\ttime\,)={b(\ttime\,)\over2\pi\tk}\sin2\pi\tk\tq\,,
\end{equation}

\noindent where the amplitude $b(\ttime\,)$ is equal to $b_i$ at the initial
time. We differentiate
equation~(120), and get
\begin{equation}
\tvv=-{db/d\ttime\over2\pi\tk}\sin2\pi\tk\tq\,,
\end{equation}

\noindent We now substitute equation~(125) into the continuity 
equation~(117), and solve for the density, according to
\begin{equation}
\trho={\tbrho\over1-b\cos2\pi\tk\tq}\,.
\end{equation}

\noindent We then substitute equations~(125) and~(126) into 
equations~(118) and~(119). 
We are left with two unknowns, $\tphi$ and~$b(\ttime\,)$. 
We eliminate~$\tphi$, and get
\begin{equation}
{d^2b\over d\ttime\,^2}=6f_nab\,.
\end{equation}

\noindent This equation is identical to equation~(104). 
The solution is therefore given by
\begin{equation}
b(\ttime\,)=A\delta_+(\ttime\,)+B\delta_-(\ttime\,)\,.
\end{equation}

\noindent By imposing the boundary condition~(123), we get
\begin{eqnarray}
A&=&b_i/\delta_+(\ttime_i)\,,\\
B&=&0\,.
\end{eqnarray}

In supercomoving variables, the solutions
for $\delta_+$ do not depend upon the cosmological parameters within
each family, except for the
case~$n=3$. The solution~(124)--(126)
describes the growth of a sinusoidal perturbation that forms a ``pancake,''
that is, a caustic
surface of infinite density, when the amplitude~$b(\ttime\,)$
reaches unity. The solution is independent of the cosmological parameters
within each family. When~$b(\ttime\,)$ reaches unity, this solution breaks 
down. At this point, strong shock waves form on both sides of the pancake,
and the collisionless approximation ($\tp=0$) is not valid anymore. 
Notice that the fluid equations are exactly the same in supercomoving variables
as in noncosmological variables. This implies that the jump conditions in 
supercomoving variables do not depend upon the cosmological parameters. Since
the solutions before shock formation do not depend upon these parameters
either, we conclude that the postshock solution will also be 
independent of the cosmological parameters within each
family. This means, in particular, that the pancake collapse problem, in the
absence of non-adiabatic physical processes like radiative cooling, 
has one solution only, when expressed in supercomoving variables.
Hence, for pancakes occurring in different cosmological models, the
values and spatial variations of any fluid variable at any time are
exactly the same as long as length is expressed in units of
the pancake wavelength while time is expressed in terms of the cosmic
scale factor $a/a_c$, where $a_c$ is the scale factor at which $b(\ttime\,)=1$
and caustics are predicted to occur as described above, independent of the 
cosmological model. Approximate analytical solutions of the
pancake collapse problem both before and after shock formation
will be presented in a forthcoming paper
(Shapiro, Struck-Marcell, \&~Martel 1998).

\section{THE ZEL'DOVICH APPROXIMATION}

Zel'dovich (1970) presented an analytical approximation for 
the growth of density perturbations,
which extrapolates the linear solution into the nonlinear regime.
This approximation is exact in the linear regime and, for locally
1D motion, remains valid into the nonlinear regime, up to
the formation of the first caustics. In this section, we derive the
supercomoving form of the Zel'dovich approximation.

Let $\rbf(t)$ be the proper position of a mass element at proper time $t$,
and let $\rbf_i=\rbf(t_i)$ be the position of that mass
element at some initial time $t_i$, in the absence of perturbation.
It can be shown that, at early time, the proper position $\rbf$ is given by
\begin{equation}
\rbf(t)=a(t)\Big[\qbf+\delta_+(\ttime\,)\pbf(\qbf)\Big]\,,
\end{equation}

\noindent where $\qbf\equiv\rbf_i/a(t_i)$, and $\pbf$ is a vector field that
is function of $\qbf$ only (not $t$), and characterizes the perturbation.
The quantity $\partial p_j/\partial q_k$ is sometimes referred to as
the deformation tensor $D_{jk}$.
The time-independence of $\pbf$ implies that at early times mass elements
are moving on straight lines. The Zel'dovich ``approximation'' consists
of assuming that equation~(131) is valid, not only at early times, but up to
the point where mass element trajectories cross, at which point the density
diverges. Writing equation~(131) in supercomoving form yields
\begin{equation}
\tr(\ttime\,)=\tqbf+\delta_+(\ttime\,)\tpbf(\tqbf)\,.
\end{equation}

\noindent 
Notice that $\tqbf$ and $\tpbf$ were computed using equation~(15) {\it without}
the factor of $a$, since $\qbf$ and $\pbf$ in equation~(131) are already
expressed in comoving variables.
In supercomoving variables, $\tqbf=[a/a(\ttime_i)]\tr_i$,
and the deformation tensor is unchanged, $\tilde D_{jk}=D_{jk}$.
As in the case of the 
collisionless pancake problem, we get a solution whose dependence upon the
cosmological model is entirely contained in the growing mode $\delta_+$.
Therefore, the solutions are identical within any given family.

\section{SUMMARY AND CONCLUSION}

In this paper, we described a set of dimensionless variables, which we call
{\it supercomoving variables}, for the description of the state and evolution
of matter in a cosmologically expanding Friedmann universe in the
Newtonian approximation (i.e. nonrelativistic gas on subhorizon scales).
These supercomoving variables have the following properties:
(1) Thermodynamic variables which depend in general upon space and time
are stationary in the absence of perturbation, so that the effects
of universal adiabatic cosmic expansion are accounted for implicitly
and do not appear explicitly as sources of time-dependence, as long as the
ratio of specific heats is $\gamma=5/3$. 
(2) Velocities reflect departures from uniform Hubble flow only and in
the absence of peculiar gravity or pressure gradients, are stationary.
(3) One advantage of these variables over standard comoving variables is that,
for ideal gas with ratio of specific heats, $\gamma=5/3$, the
fluid conservation equations in supercomoving variables are identical
to the Newtonian fluid conservation equations for a noncosmological gas,
while in standard comoving variables, extra terms appear to describe the
effects of cosmology.
(4) Similarly, the collisionless Boltzmann equation in supercomoving
variables is identical to that for a nonrelativistic, noncosmological,
collisionless gas. (5) The choice of nondimensionalizing units removes
explicit dependence of the variables and the equations which describe their
evolution including the Poisson equation for peculiar gravitational
field, on the Hubble constant $H_0$. 

Our main objective in deriving supercomoving variables was to obtain a set of 
cosmological fluid equations that resemble as closely as possible the fluid 
equations in noncosmological 
variables, and do not show any explicit dependences
upon cosmological parameters such as~$H_0$, $\Omo$, $\Omx$, or~$n$.
This was first achieved for matter-dominated models by Shandarin (S80)
for adiabatic gas, with 
terms for nonadiabatic physical processes added by 
Shapiro et al. (1983) and Shapiro \& Struck-Marcell
(1985). Here we have extended this approach to include models with a
generalized, smooth background energy density (such as a cosmological constant)
as well as additional physical processes, like viscosity and magnetic fields.
The continuity
equation, the momentum equation, and the equation of state have exactly the
same form in noncosmological and supercomoving variables. The  
cosmological drag term, that is present in the momentum equation in
comoving variables, is absent in supercomoving variables. The energy equation
has the same form in noncosmological and supercomoving variables, except 
for an additional
drag term (which is also present in comoving variables). However, unlike
the case of standard comoving variables,
that drag term vanishes for the physically interesting case~$\gamma=5/3$. Only
the Poisson equation remains very different from its noncosmological
form. In supercomoving variables, as in
comoving variables, the source of the gravitational potential is not the 
density, but rather the density contrast. 

The supercomoving variables reveal the existence of similarities among various
cosmological models. This was shown in S80 for matter-dominated models.
In comoving variables, the fluid equations, and their solutions, depend
explicitly upon the density parameter~$\Omo$. In supercomoving variables, all
matter-dominated models fall into three categories, called {\it families},
defined by~$\Omo<1$, $\Omo=1$, and~$\Omo>1$. Within each family, the
fluid equations and their solutions are independent of~$\Omo$. Therefore, using
supercomoving variables effectively reduces the number of matter-dominated
models from infinity to 3.
The existence of these families has direct practical significance for the
solution of problems involving gas dynamics and gravitational clustering
in cosmology, as follows. Any solution of the differential equations
in supercomoving variables obtained for a given initial or boundary value
problem for a particular set of values of the cosmological parameters
$H_0$ and $\Omega_0$ for which those initial and boundary values have also
been expressed in supercomoving variables immediately yields the solution for
any other values of $H_0$ and $\Omega_0$, within the same family as well,
as long as the supercomoving initial and boundary values are also the same.
This ``Family'' membership, then implies a ``similarity'' solution for
initial and boundary value problems, where the supercomoving variables are the
similarity variables.

In this paper, we generalized supercomoving variables to cosmological
models with two components, the usual nonrelativistic component, and
a uniform component,
called the X~component, whose energy density varies as~$a^{-n}$.
This generic two-component model includes as special cases several
physically motivated cosmological models, such as models with a nonzero 
cosmological constant, domain walls, cosmic strings, massive neutrinos, or
a radiation background. Each of these model is a two-parameter model, defined
by the density parameters~$\Omo$ and~$\Omx$ of the nonrelativistic and 
X~components, respectively. 

We extended the concept of family introduced in S80 for matter-dominated
models. We showed the existence of similarities among models with different
values of~$\Omo$ and~$\Omx$. For any particular value of~$n$, the various
families are defined by the value of the parameter~$\kappa$ appearing
in equation~(71). Hence, each combination~$(n,\kappa)$ corresponds to
a particular family, and, as we showed, the fluid equations and their
solutions do not depend upon the cosmological parameters~$\Omo$ and~$\Omx$
within a particular family. While Shandarin (S80) reduced a one-parameter
model (the parameter being~$\Omo$) to three no-parameter models, we are 
reducing two-parameter models (one for each value of~$n$, the parameters
being~$\Omo$ and~$\Omx$) to one-parameters models (the parameter 
being~$\kappa$). Therefore, as in the case of matter-dominated models, 
solutions obtained in supercomoving 
variables for a particular model (that is, for 
particular values of~$\Omo$ and~$\Omx$) can be applied to other models
as long as these models are in the same family. In this paper, we focused 
our attention on 14 particular families which are physically interesting:
(1)~the Einstein-de Sitter model, (2)~open, matter-dominated models,
(3)~closed, matter-dominated models, (4)~flat models with a nonclumping
background of nonrelativistic matter, such as would describe the case of
massive neutrinos in the small wavelength limit, 
(5)~open models with a nonrelativistic matter background, such as
massive neutrinos, (6)~closed models with a nonrelativistic matter
background, such as
massive neutrinos, (7)~marginally bound models with cosmic strings,
\footnote{There is an important distinction, often forgotten, between
the terms ``closed'' and ``bound.'' The former refers to the curvature of
the universe, and is determined by the sign of~$\Omo+\Omx-1$; the latter
refers to whether the universe will expand forever or eventually recollapse.
For matter dominated models (and also models with massive neutrinos), closed
models are always bound and open models are always unbound. But this is not
true in general for other models.} (8)~unbound models with cosmic strings,
(9)~bound models with cosmic strings, (10)~flat models with a radiation
background, (11)~flat models with a nonzero cosmological constant,
(12)~critical models (i.e. asymptotically static)
with a nonzero cosmological constant, (13)~flat models
with domain walls, and (14)~critical models 
(i.e. asymptotically static) with domain walls. Table~4
summarizes the values of the parameters that define these various families.

To illustrate the use of supercomoving variables and the value of the
``family'' membership they reveal, we have,
for each of these families, described the linear growth of density fluctuations
in supercomoving variables,
the supercomoving Jeans length, and the nonlinear growth of 1D planar density
fluctuations -- the cosmological pancake problem -- prior
to caustic formation. 
The solutions of these problems are independent of the particular values
of the cosmological
parameters $\Omo$ and~$\Omx$ within a given family, with the exception of 
the models with a nonrelativistic matter background such as
massive neutrinos, for which the solutions must remain 
explicitly parametrized by~$\Omo$
and~$\Omx$. The model with massive neutrinos differs from
the other models in that the energy density of the two components both vary
as~$a^{-3}$. As a result, there is no characteristic time scale when the    
universe goes from being matter-dominated to being X~component-dominated, 
or vice-versa. This absence of characteristic time scale implies that 
solutions for different models are genuinely different, and cannot
be turned into one another by a scaling transformation, as for the other
models. The only exception is the evolution of the scale factor (eq.~[76]),
which is made independent of~$\Omo$ and~$\Omx$ by our introduction of
the parameter~$f_n$ in the definition
of the fiducial time~$t_\ast$ (eq.~[25]). 
By this choice, however, the form of the Poisson
equation is different for $n=3$ models. 

Finally, we have shown how the Navier-Stokes equation and the viscous
energy equation are expressed in supercomoving variables, and presented
supercomoving forms for the inviscid equation for vorticity evolution
and the MHD equations, including the Biermann battery term. The latter
two make clear how the vorticity and cosmic magnetic fields generated
by large-scale structure formation, as discussed recently by Kulsrud et al.
(1997), are related to each other and how they scale with characteristic 
time and length scales for cosmic structure formation in most cosmological
models of interest.

\acknowledgments
This work was supported by NASA Astrophysics Theory Grant Program
Grant NAG5-2785 and NSF Grant ASC9504046.

%

%

\appendix
\section{ADDITIONAL PHYSICAL PROCESSES}

\subsection{Heating, Cooling, and Thermal Conduction}

Shapiro et al. (1983), and Shapiro \&~Struck-Marcell (1985) 
showed how to transform the terms for the additional physical processes
of radiative cooling, external heating, and thermal conduction
into supercomoving variables. This requires only a
modification of the energy equation. In noncosmological 
variables, this equation becomes
\begin{equation}
{\partial\uth\over\partial t} +\ubf\cdot\grad\uth=-{p\over\rho}\divg\ubf
+{\Gamma-\Lambda\over\rho}-{\divg\Sbf\over\rho}\,,
\end{equation}

\noindent where $\Gamma$ is the volume heating rate, 
$\Lambda$ is the volume cooling rate, and $\Sbf$ is the conductive flux.
We define the supercomoving heating rate~$\tilde\Gamma$,
cooling rate~$\tilde\Lambda$, and conductive flux~$\tilde\Sbf$ as
\begin{eqnarray}
\tilde\Gamma&&={a^7\Gamma\over\Gamma_\ast}\,,\\
\tilde\Lambda&&={a^7\Lambda\over\Lambda_\ast}\,,\\
\tilde\Sbf&&={a^6\Sbf\over S_\ast}\,,
\end{eqnarray}

\noindent where
\begin{eqnarray}
\Gamma_\ast&=&\Lambda_\ast\equiv{p_\ast\over t_\ast}\,,\\
S_\ast&\equiv&\rho_\ast v_\ast^3\,.
\end{eqnarray}

\noindent Equation~(A1) becomes, in supercomoving variables,
\begin{equation}
{\partial\tuth\over\partial\ttime} +\tv\cdot\tgrad\tuth=-{\cal H}
(3\gamma-5)\tuth-{\tp\over\trho}\tgrad\cdot\tv
+{\tilde\Gamma-\tilde\Lambda\over\trho}
-{\tgrad\cdot\tilde\Sbf\over\trho}\,.
\end{equation}

\noindent This equation has the same form as equation~(A1), except for the
drag term, which vanishes for $\gamma=5/3$.

\subsection{Viscosity}

If the fluid has a finite viscosity, the momentum equation is replaced
by the Navier-Stokes equation,
\begin{equation}
{\partial\ubf\over\partial t} 
+(\ubf\cdot\grad)\ubf=-\grad\Phi-{\grad p\over\rho}
+{\eta\over\rho}\nabla^2\ubf+{\zeta+\eta/3\over\rho}\grad(\divg\ubf)\,,
\end{equation}

\noindent (Landau \& Lifshitz 1987)
where $\eta$ and $\zeta$ are the coefficients of viscosity.
The presence of viscosity dissipates kinetic energy into heat, hence the
energy equation must be modified accordingly. Its final form is
\begin{equation}
{\partial\uth\over\partial t} +\ubf\cdot\grad\uth=-{p\over\rho}\divg\ubf
+{1\over\rho}\sigma_{ij}'{\partial u_i\over\partial r_j} \,,
\end{equation}

\noindent
where the stress tensor~$\sigma'$ is defined by
\begin{equation}
\sigma_{ij}'=\eta\biggl({\partial u_i\over\partial r_j} 
+{\partial u_j\over\partial r_i}
-{2\over3}\delta_{ij}{\partial u_k\over\partial r_k} \biggr)+\zeta\delta_{ij}
{\partial u_k\over\partial r_k}
\,.
\end{equation}

\noindent
We define the following supercomoving coefficients of viscosity and 
stress tensor:
\begin{eqnarray}
\tilde\eta&=&{a^3\eta\over\eta_\ast}\,,\\
\tilde\zeta&=&{a^3\zeta\over\zeta_\ast}\,,\\
\tilde\sigma_{ij}'&=&\tilde\eta\biggl({\partial\tvv_i\over\partial\trr_j} 
+{\partial\tvv_j\over\partial\trr_i}
-{2\over3}\delta_{ij}{\partial\tvv_k\over\partial\trr_k} \biggr)
+\tilde\zeta\delta_{ij}\biggl({\partial\tvv_k\over\partial\trr_k}
+6{\cal H}\biggr)
\,,
\end{eqnarray}

\noindent where
\begin{equation}
\eta_\ast=\zeta_\ast\equiv{\rho_\ast r_\ast^2\over t_\ast}\,.
\end{equation}

\noindent With these definitions, equations~(A8) and~(A9) become,
in supercomoving variables,
\begin{eqnarray}
&&{\partial\tv\over\partial\ttime} 
+(\tv\cdot\tgrad)\tv=-\tgrad\tphi-{\tgrad\tp\over\trho}
+{\tilde\eta\over\trho}\tilde\nabla^2\tv
+{\tilde\zeta+\tilde\eta/3\over\trho}\tgrad(\tgrad\cdot\tv)\,,\\
&&{\partial\tuth\over\partial\ttime} +\tv\cdot\tgrad\tuth
=-{\cal H}(3\gamma-5)\tuth
-{\tp\over\trho}\tgrad\cdot\tv+{1\over\trho}\tilde\sigma_{ij}'
{\partial\tv_i\over\partial\tx_j} 
+{9\tilde\zeta\over\trho}{\cal H}^2
\,.
\end{eqnarray}

\subsection{Vorticity}

The vorticity is defined by
\begin{equation}
\omegabf=\nabla\times\ubf\,.
\end{equation}

\noindent In the absence of viscosity, the vorticity evolves according to
\begin{equation}
{\partial\omegabf\over\partial t}+\nabla\times(\omegabf\times\ubf)
=-\nabla p\times\nabla\biggl({1\over\rho}\biggr)\,.
\end{equation}

\noindent To convert these equations to supercomoving form, we define
\begin{equation}
\tomega={a^2\omegabf\over\omega_\ast}\,,
\end{equation}

\noindent where
\begin{equation}
\omega_\ast\equiv{1\over t_\ast}\,.
\end{equation}

\noindent Equations~(A17) and~(A18) become, in supercomoving variables,
\begin{equation}
\tomega=\tgrad\times\tv\,,
\end{equation}

\noindent and
\begin{equation}
{\partial\tomega\over\partial\ttime}+\tgrad\times(\tomega\times\tv)
=-\tgrad\tp\times\tgrad\biggl({1\over\trho}\biggr)\,.
\end{equation}

\subsection{Magnetic Fields}

In the
presence of magnetic fields, the momentum equation must be modified to take
into account the magnetic tension and
pressure. In the MHD approximation (where
the fluid is assumed to have infinite conductivity), the momentum equation
becomes
\begin{equation}
{\partial\ubf\over\partial t} 
+(\ubf\cdot\grad)\ubf=-\grad\Phi-{\grad p\over\rho}
+{(\curl\Bbf)\times\Bbf\over4\pi\rho}\,,
\end{equation}

\noindent where~$\Bbf$ is the magnetic field. The time evolution of
the magnetic field is described by the following equation,
\begin{equation}
{\partial\Bbf\over\partial t}=\curl(\ubf\times\Bbf)
+{c\over n_e^2e}\grad p_e\times\grad n_e\,.
\end{equation}

\noindent This is a standard MHD equation, except for the last term,
which we have added to take into account the possibility of magnetic field
generation in an ionized gas
by a process known as the Biermann battery (Biermann 1950).
In this term, $p_e$ and $n_e$ are the electron pressure and
number density, respectively. If we assume that the ionized fraction $\chi$
is uniform and the electron, ion, and neutral temperatures are equal,
this equation reduces to
\begin{equation}
{\partial\Bbf\over\partial t}=\curl(\ubf\times\Bbf)
-{cm_{\rm H}\over e(1+\chi)}\grad p\times\grad\biggl({1\over\rho}\biggr)\,,
\end{equation}

\noindent (Kulsrud et al. 1997),
where $m_{\rm H}$ is the hydrogen mass.
To convert these equations to supercomoving variables, we consider two
different definitions for the supercomoving magnetic field, 
\begin{eqnarray}
\tilde\Bbf_1&=&{a^{5/2}\Bbf\over B_\ast}\,,\\
\tilde\Bbf_2&=&{a^2\Bbf\over B_\ast}\,,
\end{eqnarray}

\noindent where
\begin{equation}
B_\ast\equiv\rho_\ast^{1/2}v_\ast\,.
\end{equation}

\noindent Using the definition~(A26),
the momentum equation becomes, in supercomoving variables,
\begin{equation}
{\partial\tv\over\partial\ttime} 
+(\tv\cdot\tgrad)\tv=-{\tgrad\tp\over\trho}-\tgrad\tphi
+{(\tgrad\times\tilde\Bbf_1)\times\tilde\Bbf_1\over4\pi\trho}\,,
\end{equation}

\noindent which has the same form as equation~(A23), and
equation~(A24) becomes
\begin{equation}
{\partial\tilde\Bbf_1\over\partial\ttime}-{{\cal H}\over2}
\tilde\Bbf_1=\tgrad\times(\tv\times\tilde\Bbf_1)
-{a^{1/2}\over\rho_\ast^{1/2}r_\ast}{cm_{\rm H}\over e(1+\chi)}
\tgrad\tp\times\tgrad\biggl({1\over\trho}\biggr)\,.
\end{equation}

\noindent Comparing equations~(A24) and~(A30), we see that the conversion to
supercomoving variables introduces an ``antidrag'' term 
${\cal H}\Bbf/2$ (which also appears in standard comoving 
variables, though with a different coefficient), and an extra factor of
$a^{1/2}$ in the Biermann term. Using the second definition~(A27), instead,
the momentum equation becomes
\begin{equation}
{\partial\tv\over\partial\ttime} 
+(\tv\cdot\tgrad)\tv=-{\tgrad\tp\over\trho}-\tgrad\tphi
+{a(\tgrad\times\tilde\Bbf_2)\times\tilde\Bbf_2\over4\pi\trho}\,,
\end{equation}

\noindent and equation~(A24) becomes
\begin{equation}
{\partial\tilde\Bbf_2\over\partial\ttime}
=\tgrad\times(\tv\times\tilde\Bbf_2)
-{1\over\rho_\ast^{1/2}r_\ast}{cm_{\rm H}\over e(1+\chi)}
\tgrad\tp\times\tgrad\biggl({1\over\trho}\biggr)\,,
\end{equation}

\noindent which has a form very similar to equation~(A24).

As noted recently by Kulsrud et al. (1997), there is a close correspondence 
between the inviscid equation for vorticity evolution, equation~(A18) and
the MHD equation for $\Bbf$ in the presence of the Biermann battery term, 
equation~(A25), especially when $\Bbf$ is expressed in the same units as 
vorticity, by replacing $\Bbf$ by the cyclotron frequency 
$\omegabf_{\rm cyc}=e\Bbf/m_{\rm H}c$. 
In the case in which both vorticity and magnetic
field are initially zero, this leads to the remarkably simple result that
\begin{equation}
\omegabf_{\rm cyc}=-{\omegabf\over1+\chi}\,.
\end{equation}

We can write equation~(A32) in terms of $\tomega_{\rm cyc}$ defined by
\begin{equation}
\tomega_{\rm cyc}={\omegabf_{\rm cyc}a^2\over\omega_{\rm cyc\ast}}\,,
\end{equation}

\noindent where
\begin{equation}
\omega_{\rm cyc\ast}\equiv{1\over t_\ast}=\omega_\ast\,.
\end{equation}

\noindent This yields
\begin{equation}
{\partial\tomega_{\rm cyc}\over\partial\ttime}=
\tgrad\times(\tv\times\tomega_{\rm cyc})
-{1\over1+\chi}\tgrad\tp\times\tgrad\biggl({1\over\trho}\biggr)\,.
\end{equation}

\noindent A comparison of equations~(A22) and~(A36) then replaces
equation~(A33) by
\begin{equation}
\tomega_{\rm cyc}=-{\tomega\over1+\chi}\,.
\end{equation}

%

\def \ms {\llap{$-$}}

\begin{deluxetable}{ccccccc}
\footnotesize
\tablecaption{SUPERCOMOVING AGES OF THE UNIVERSE (PRESENT AND ASYMPTOTIC) 
FOR FLAT MODELS WITH $n=0$ AND $n=1$.\tablenotemark{a}}
\tablewidth0pt
\tablehead{
\colhead{$\displaystyle{\Omega_0\atop{}}$} & 
\colhead{$\displaystyle{{}\atop a_0}$} & 
\colhead{$\displaystyle{n=0\atop\tilde t_0}$} &
\colhead{$\displaystyle{{}\atop\tilde t_\infty}$} &
\colhead{$\displaystyle{{}\atop a_0}$} & 
\colhead{$\displaystyle{n=1\atop\tilde t_0}$} &
\colhead{$\displaystyle{{}\atop\tilde t_\infty}$}}
\startdata
0.01 & 4.6261 &    0.20324 & 0.21490 & 9.9499 &    0.27290 & 0.28350 \nl
0.02 & 3.6593 &    0.19630 & 0.21490 & 7.0000 &    0.26558 & 0.28350 \nl
0.05 & 2.6684 &    0.18015 & 0.21490 & 4.3589 &    0.24728 & 0.28350 \nl
0.10 & 2.0801 &    0.15834 & 0.21490 & 3.0000 &    0.22080 & 0.28350 \nl
0.20 & 1.5874 &    0.11014 & 0.21490 & 2.0000 &    0.17131 & 0.28350 \nl
0.50 & 1.0000 &    0.00000 & 0.21490 & 1.0000 &    0.00000 & 0.28350 \nl
0.80 & 0.6299 & \ms0.21178 & 0.21490 & 0.5000 & \ms0.33823 & 0.28350 \nl
0.90 & 0.4807 & \ms0.38043 & 0.21490 & 0.3333 & \ms0.63233 & 0.28350 \nl
0.99 & 0.2162 & \ms1.07573 & 0.21490 & 0.1005 & \ms2.02892 & 0.28350 \nl
\enddata
\tablenotetext{a}{Present scale factor $a_0\equiv a(\ttime_0)$, where
$\ttime_0=\ttime(t_0)$, $t_0=\hbox{present}$
age of the universe in proper time, and $\ttime_\infty=\ttime(t=\infty)$,
where $t$ is proper time. We note that $\ttime(t=0)=-\infty$ in all cases.}
\end{deluxetable}

\begin{deluxetable}{lccc}
\footnotesize
\tablecaption{SELECTED FAMILIES OF INTEREST}
\tablewidth{0pt}
\tablehead{\colhead{Family} & \colhead{$n$} & \colhead{$\kappa$} 
& \colhead{Density Parameters} }
\startdata
Einstein-de Sitter model                             
                              & NA\tablenotemark{a} 
                              & NA & $\Omo=1$, $\Omx=0$            \nl
Open, matter-dominated models                          
                              & NA & NA & $\Omo<1$, $\Omx=0$       \nl
Closed, matter-dominated models     
                              & NA & NA & $\Omo>1$, $\Omx=0$       \nl
Flat models with massive neutrinos                    
                              & 3   & NA & $\Omo+\Omx=1$           \nl
Open models with massive neutrinos                 
                              & 3   & NA & $\Omo+\Omx<1$           \nl
Closed models with massive neutrinos                  
                              & 3   & NA & $\Omo+\Omx>1$           \nl
Marginally bound models with cosmic strings         
                              & 2   & NA & $\Omo=1$, any $\Omx$    \nl
Unbound models with cosmic strings                   
                              & 2   & NA & $\Omo<1$, any $\Omx$    \nl
Bound models with cosmic strings                    
                              & 2   & NA & $\Omo>1$, any $\Omx$    \nl
Flat models with a radiation background              
                              & 4   & 0   & $\Omo+\Omx=1$          \nl
Flat models with a nonzero cosmological constant     
                              & 0   & 0   & $\Omo+\Omx=1$          \nl
Critical models with a nonzero cosmological constant 
                              & 0   & $-3/4^{1/3}$ & Equation (68) \nl 
Flat models with domain walls                       
                              & 1   & 0   & $\Omo+\Omx=1$          \nl
Critical models with domain walls                    
                              & 1   & $-2$ & Equation (68)         \nl
\enddata
\tablenotetext{a}{For entries labeled ``NA,'' this means that the parameter
is ``not applicable.''}
\end{deluxetable}

%

\clearpage
\begin{center}
Figure Captions
\end{center}

\figcaption{Evolution of Cosmological Models in Supercomoving Variables.
(a) (Top panel) Cosmic Scale Factor $a(\ttime\,)$ versus time $\ttime$
for various models.
Solid curves (from left to right): open, flat, and closed, 
matter-dominated models. The solid dot indicates the location of the
present for the flat model. Dotted curve: flat model with radiation.
This curve merges
with the solid curve for the flat matter-dominated model at late time.
Short-dashed curves: critical models with a nonzero cosmological constant
(lower curve) and with domain walls (upper curve).
Long-dashed curves: flat models with a cosmological constant and with domain 
walls. (b) (Bottom panel)
Supercomoving Hubble parameter ${\cal H}=a^{-1}da/d\ttime$
versus $\ttime$, for the
same models as plotted in~(a)}

\figcaption{Families of Cosmological Models
(a) (Top left panel) Family membership curves
for $n=0$. The curves are isocontours of the family 
membership parameter $\kappa$, as defined by 
equation~(71), for particular values of $\kappa$, as labeled.
The dashed curves represent particular cases mentioned in the text. For values
in the shaede ragion, the universe does not begin with a Big Bang. 
(b) (Top right panel) Enlargement of lower right corner
of (a). Models located below the dashed curve are bound, others are unbound.
(c) (Middle left panel) Same as (a), except for models with $n=1$. (d) (Middle
right panel) Family Memberships for $n=2$ models. There are 3 families,
represented by the dashed line, and the 2 half-planes. (e) (Bottom left
panel) Same as (d) except for models with $n=3$. (f) (Bottom right) Same as (a)
except for models with $n=4$. Shaded area in (a) and (c) represents 
combination of parameters incompatible with the existence of a Big Bang.}


\clearpage
\begin{thebibliography}{}

\bibitem{}
Albretch, A., \& Steinhardt, P. J. 1982, Phys.Rev.Letters, 48, 1220

\bibitem{}
Biermann, L. 1950, Z. Naturforsch, 5a, 65

\bibitem{}
Bildhauer, S., Buchert, T., \& Kasai, M. 1992, A\&A, 263, 23

\bibitem{}
Bond, J. R., Efstathiou, G., \& Silk, J. 1980, Phys.Rev.Letters, 45, 1980

\bibitem{}
Charlton, J. C., \& Turner, M. S. 1987, ApJ, 313, 495

\bibitem{}
Carroll, S. M., Press, W. H., \& Turner, E. L. 1992, ARA\&A, 30, 499

\bibitem{}
Dekel, A., Bertschinger, E., Yahil, A., Strauss, M. A., Davis, M., \&
Huchra, J. P. 1993, ApJ, 412, 1

\bibitem{}
Dodelson, S., Gates, E. I., \& Turner, M. S. 1996, Science,
274, 69

\bibitem{}
Doroshkievich, A., Ryabenkii, V., \& Shandarin, S. 1997, Astrophysics, 9, 144

\bibitem{}
Edwards, D., \& Heath, D. 1976, Astron.Sp.Sci., 41, 183

\bibitem{}
Freedman, W. 1996, preprint astro-ph/9612024

\bibitem{}
Fry, J. N. 1985, Phys.Letters B, 158, 211

\bibitem{}
Giovanelli, R., Haynes, M., da Costa, L., Freudling, W., Salzer, J.,
\& Wegner, G. 1996, preprint astro-ph/9612072

\bibitem{}
Groth, E. J., \& Peebles, P. J. E. 1975, A\&A, 41, 143

\bibitem{}
Guth, A. H. 1981, Phys.Rev.D, 23, 347

\bibitem{}
Krauss, L. M., \& Turner, M. S. 1995, Gen.Relativ.Grav., 27, 1137

\bibitem{}
Kulsrud, R. M., Cen, R., Ostriker, J. P., \& Ryu, D., 1997,
ApJ, 480, 481

\bibitem{}
Lahav, O., Lilje, P. B., Primack, J. R., \& Rees, M. J. 1991, MNRAS, 251, 128

\bibitem{}
Landau, L. D., \& Lifshitz, E. M. 1987, Fluid Mechanics (New York:
Pergamon Press), pp.~44--46

\bibitem{}
Linde, A. 1982, Phys.Letters B, 108, 389

\bibitem{}
Lyth, D. H., \& Steward, E. D. 1990, Phys.Letters B, 252, 336

\bibitem{}
Martel, H. 1991, ApJ, 377, 7

\bibitem{}
Martel, H 1995, ApJ, 445, 537

\bibitem{}
M\'esz\'aros, P. 1974, A\&A, 37, 225

\bibitem{}
Ostriker, J. P. 1993, ARA\&A, 31, 689

\bibitem{}
Ostriker, J. P., \& Steinhardt, P. J. 1995, Nature, 377, 600

\bibitem{}
Peebles, P. J. E. 1980, The Large-Scale Structure of the Universe
(Princeton: Princeton University Press) (P80)

\bibitem{}
Peebles, P. J. E. 1993, Principles of Physical Cosmology 
(Princeton: Princeton University Press)

\bibitem{}
Ratra, B., \& Peebles, P. J. E. 1994, ApJ, 432, L5

\bibitem{}
Riess, A., Kirshner, R. P., \& Press, W. 1995, ApJ, 438, L17

\bibitem{}
Shandarin, S. F. 1980, Astrofizika, 16, 769 (S80)

\bibitem{}
Shapiro, P. R., Giroux, M. L., \& Babul, A. 1994, ApJ, 427, 25

\bibitem{}
Shapiro, P. R., Martel, H., Villumsen, J. V., \& Owen, J. M. 1996,
ApJ Suppl, 103, 269

\bibitem{}
Shapiro, P. R., Struck-Marcell, C., \& Melott, A. L. 1983, ApJ, 275, 413

\bibitem{}
Shapiro, P. R., \& Struck-Marcell, C. 1985, ApJ Suppl, 57, 205

\bibitem{}
Shapiro, P. R., Struck-Marcell, C., \& Martel, H. 1998, in preparation

\bibitem{}
Silveira, V., \& Waga, I. 1994, Phys.Rev.D, 50, 4890

\bibitem{}
Sunyaev, R. A., \& Zel'dovich, Ya. B. 1972, A\&A, 20, 189

\bibitem{}
Tonry, J. L. 1993, in Relativistic Astrophysics and Cosmology,
Ann. NY Ac. Sci., 688, 113

\bibitem{}
Turner, M. S., \& White, M. 1997, preprint astro-ph/9701138

\bibitem{}
Valinia, A., Shapiro, P. R., Martel, H., \& Vishniac, E. T. 1997, ApJ, 479, 46

\bibitem{}
Voit, G. M. 1996, ApJ, 465, 548

\bibitem{}
Wasserman, I. 1981, ApJ, 248, 1

\bibitem{}
Weinberg, S. 1972, Gravitation and Cosmology (New York: Wiley)

\bibitem{}
Zel'dovich, Ya. B. 1970, A\&A, 5, 84

\end{thebibliography}
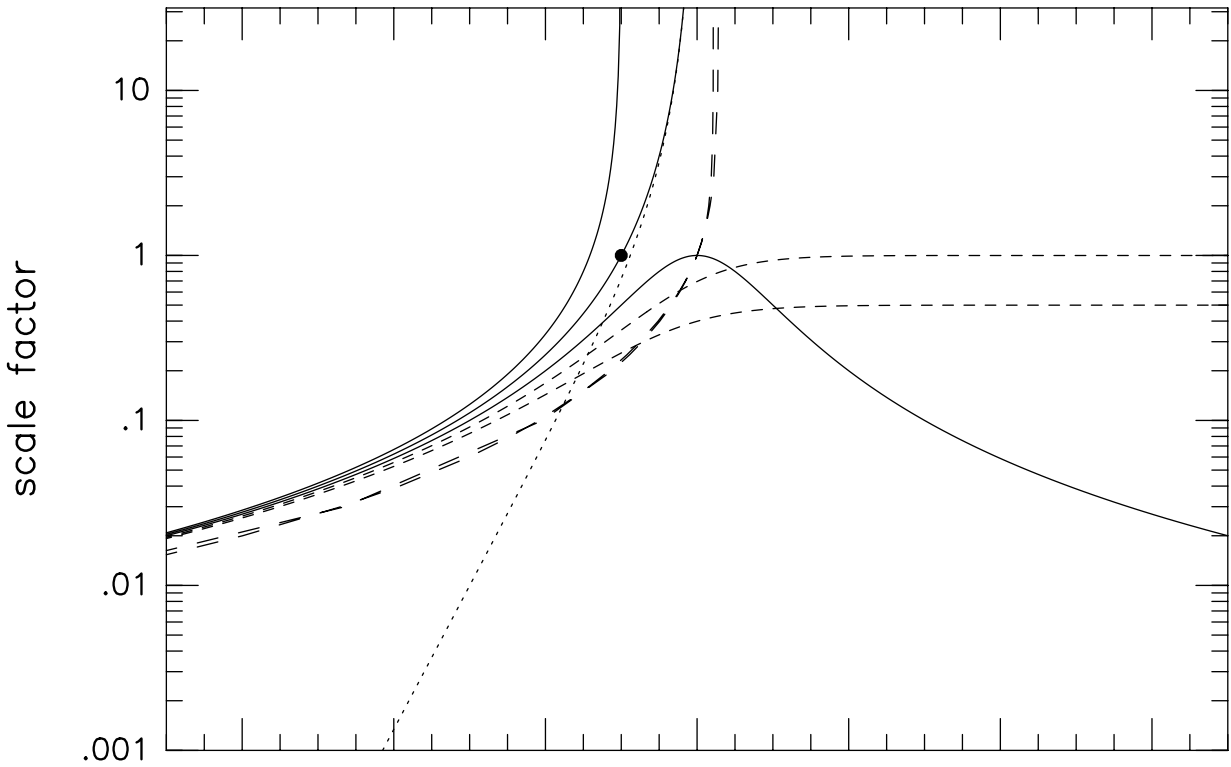
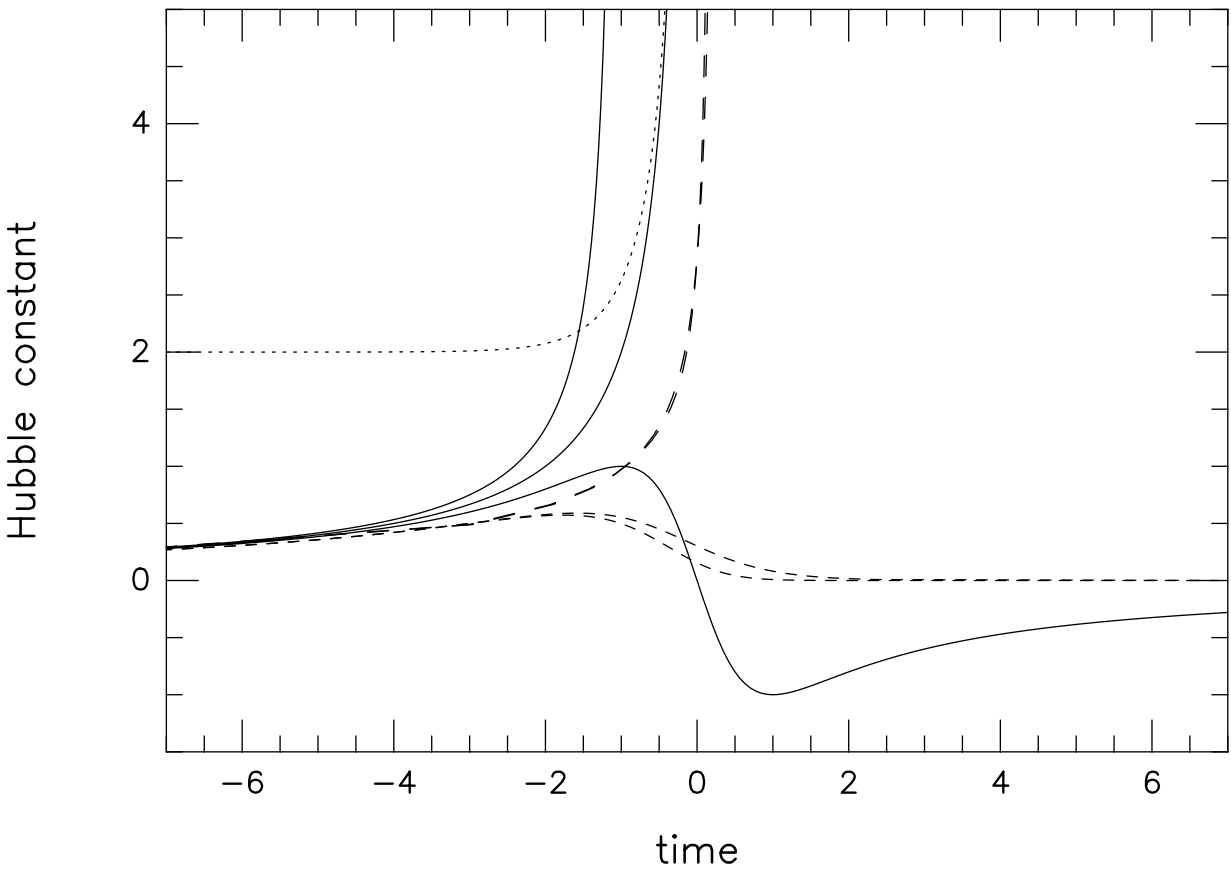
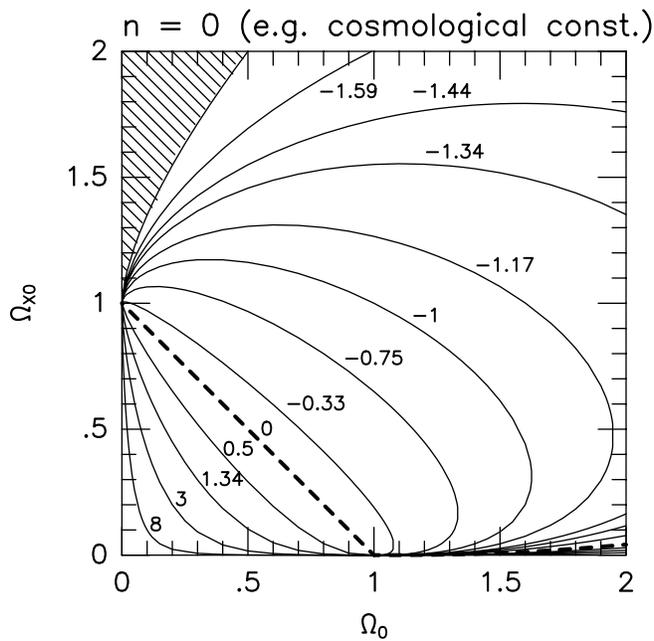
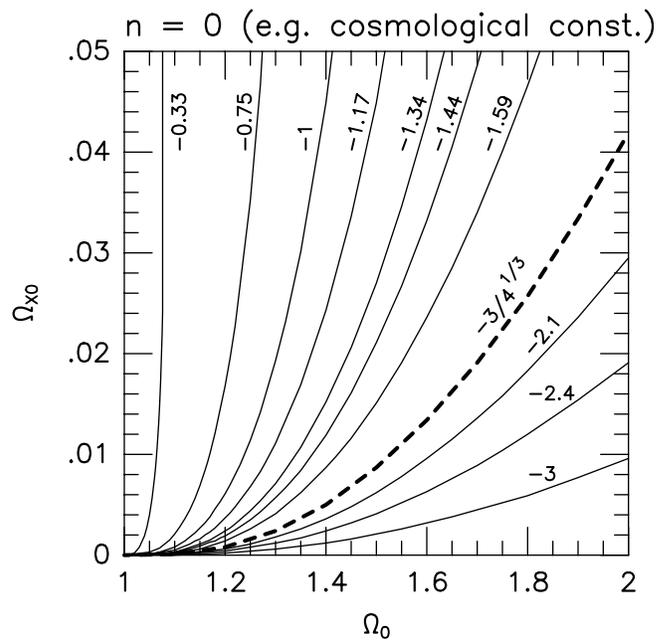
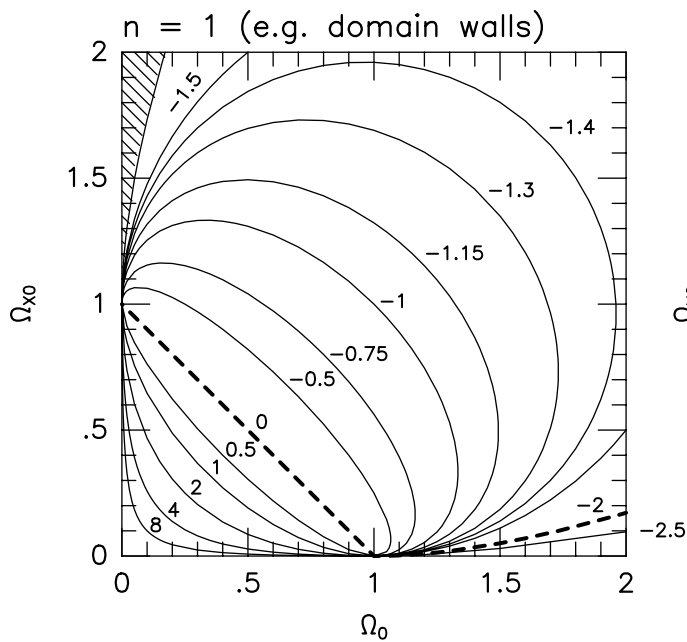
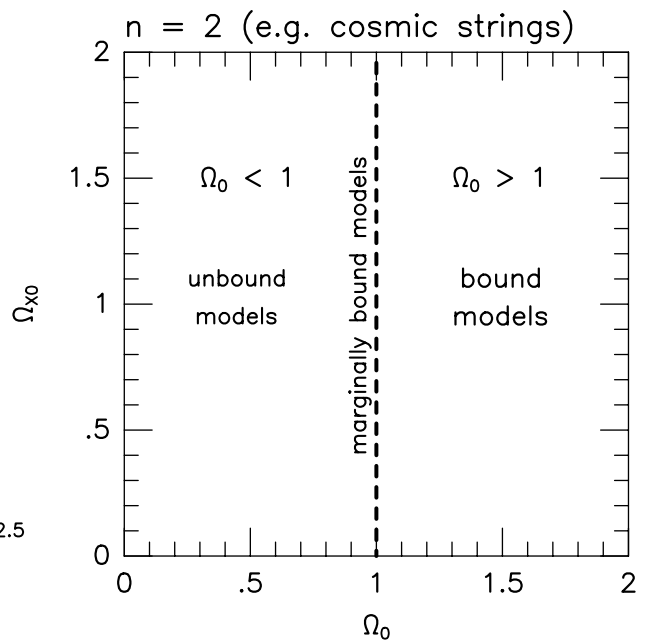
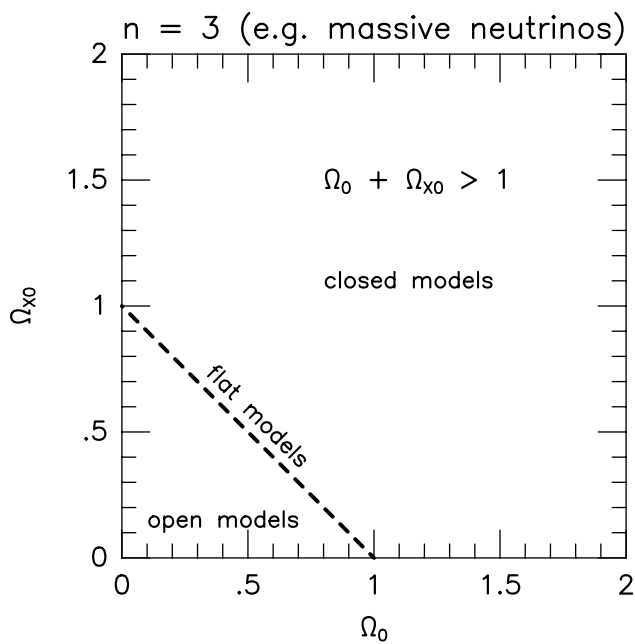
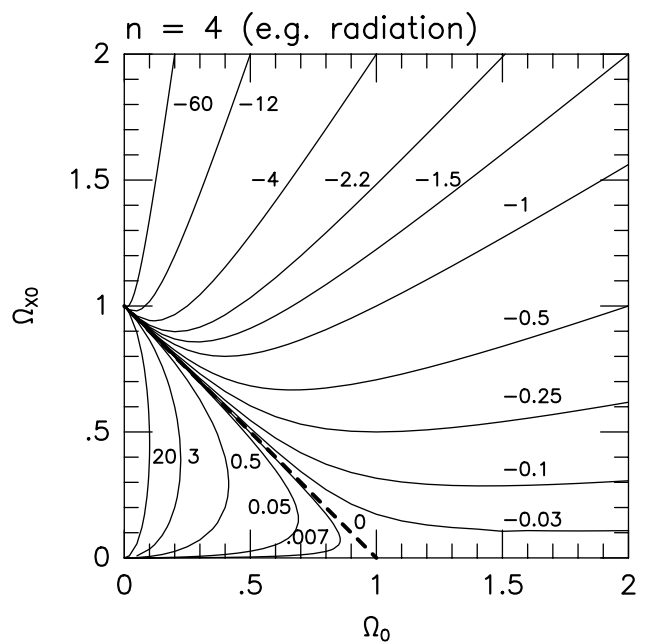
\end{document}